\documentclass[10pt,twocolumn,pra,superscriptaddress,aps]{revtex4-2}       
\usepackage{graphicx,subfigure}                           
\usepackage[normalem]{ulem}                               
\usepackage[dvipsnames]{xcolor}                           
\usepackage{physics}                                      
\usepackage{dsfont}                                       
\usepackage{amsmath}									  
\definecolor{mycolor}{rgb}{0.188, 0.196, 0.654}           
\definecolor{indiagreen}{rgb}{0.07, 0.53, 0.03}
\definecolor{indigo}{rgb}{0.29, 0.0, 0.51}
\usepackage[
colorlinks=true, 
linkcolor=mycolor, 
citecolor=mycolor, 
urlcolor=mycolor]{hyperref}

\newcommand{\Coin}[1]{$\mathcal{C}_{#1}$}

\newcommand{\Red}[1]{\textcolor{red}{#1}}
\newcommand{\Blue}[1]{\textcolor{blue}{#1}}

\begin{document}
	
\title{Parrondo's paradox in quantum walks with inhomogeneous coins}
\author{Vikash Mittal}
    \email{vikashmittal.iiser@gmail.com}
    \affiliation{Department of Physics, National Tsing Hua University, Hsinchu 30013, Taiwan}

\author{Yi-Ping Huang}
    \email{yphuang@phys.nthu.edu.tw} 
    \affiliation{Department of Physics, National Tsing Hua University, Hsinchu 30013, Taiwan}
    \affiliation{Physics Division, National Center for Theoretical Sciences, Taipei 10617, Taiwan}
    \affiliation{Institute of Physics, Academia Sinica, Taipei 115, Taiwan}

\begin{abstract}
    Parrondo's paradox, a counterintuitive phenomenon where two losing strategies combine to produce a winning outcome, has been a subject of interest across various scientific fields, including quantum mechanics. In this study, we investigate the manifestation of Parrondo's paradox in discrete-time quantum walks. We demonstrate the existence of Parrondo's paradox using site- and time-dependent coins without the need for a higher-dimensional coin or adding decoherence to the system. Our results enhance the feasibility of practical implementations and provide deeper insights into the underlying quantum dynamics, specifically the propagation constrained by the interference pattern of quantum walks. The implications of our results suggest the potential for more accessible and efficient designs in quantum transport, broadening the scope and application of Parrondo's paradox beyond conventional frameworks.
\end{abstract}

\maketitle

\section{Introduction}
Parrondo's paradox~\cite{Abbott1999,Abbott1999a,Abbott2000,Parrondo2000,Parrondo2000a}, a counterintuitive phenomenon where two losing strategies combine to produce a winning outcome, has been an active topic of research in physics and the mathematics community~\cite{Harmer2002,Abbott2010,Cheong2019,Lai2020}. Initially, it was conceptualized in classical game theory~\cite{Parrondo2000a, Parrondo2004,Dinis2008}, but this paradox has found applications across various fields, including finance~\cite{Tamarkin2005,Ma2019}, evolutionary biology~\cite{cheong_paradoxical_2016, koh_generalized_2020,tan_nomadic-colonial_2017,tan_predator_2019}, and statistical physics~\cite{Chen2010,cheung_winning_2016,Flitney2002,flitney_quantum_2003,Dinis2008,Kocarev2002,Ye2013,fotoohinasab_denoising_2018,meyer_quantum_1996,Kay2003,miszczak_constructing_2022,jia_parrondo_2020,ejlali_parrondos_2020,ye_effects_2021}. The search for Parrondo's paradox has recently extended into the domain of quantum mechanics, particularly within the framework of quantum walks~\cite{Meyer2002,meyerQUANTUMPARRONDOGAMES2002,Meyer2003,Flitney2004, Kosik2007,Chandrashekar2011,Flitney2012,Guo2013,Rajendran2018,Rajendran2018a,Machida2018,Lai2020,Lai_PRE_2020a,Duarte2020, Abbott2020,Bauer_PRE_2021,Trautmann_2022,Bauer_PRE_2022,Panda2022,Bauer_PRE_2023,Mielke2023,Jan2023, Ximenes2024}. The first (and only as per our knowledge) experimental implementation of Parrondo's paradox in quantum walks was presented in Ref.~\cite{Abbott2020} using an optical setup.

Quantum walks, the quantum analog of classical random walks~\cite{Aharonov1993, Kempe2003}, have proven to be a powerful tool for modeling and analyzing quantum systems~\cite{Ambainis2003, Childs2004, Shenvi2003, Agliari2010}. Unlike classical random walks, quantum walks are governed by quantum interference and superposition principles, leading to distinct behaviors such as faster mixing and hitting times~\cite{VenegasAndraca2012}. These quantum effects offer deeper insights into the propagation of quantum states and their underlying dynamics, with potential application in quantum computation and quantum information processing~\cite{Nicola2014,Childs2009,Childs2013,Lovett2010,QWMeasurement2013,QWMeasurement2023,Mittal2023}. In particular, discrete-time quantum walks (DTQW) offer a versatile platform for studying complex quantum phenomena, such as topological phases~\cite{Kitagawa2010,Kitagawa2012,Asboth2012,Asboth2013,Asboth2015,Mittal2021} where the walker moves in discrete steps determined by a quantum coin.

In this paper, we investigate the manifestation of Parrondo's paradox in discrete-time quantum walks. Earlier studies have demonstrated Parrondo's paradox in quantum systems using various approaches. For instance, the authors in Ref.~\cite{Rajendran2018} have employed a three-state coin, and in Ref.~\cite{Lai_PRE_2020} a four-state coin to induce the paradox, which are examples of utilizing a higher-dimensional coin to realize the paradox. While in Ref.~\cite{Bauer_PRE_2023}, the authors have introduced decoherence to achieve similar outcomes. In Ref.~\cite{Duarte_PRE_2020}, the authors have realized Parrondo's paradox with time-dependent coins, although in a different setting than ours, where the coin operator is chosen to be a Hadamard coin and time dependence is encoded linearly in the coin. Although these methods have successfully shown the paradox in quantum walks, they add layers of complexity that may hinder practical implementation.

However, here we report our findings, which reveal that neither of these conditions are necessary. By employing site- and time-dependent coins, we demonstrate the existence of Parrondo's paradox in a simpler and more elegant manner. In a DTQW, a quantum particle moves across a lattice governed by two key operators: the coin operator and the translation operator. The coin operator acts on the internal state of the particle (often called the coin state), creating a superposition that determines the probability of the particle moving in different directions. The translation operator then shifts the particle along the lattice based on the internal state of the particle. In conventional quantum walks, the coin operator is independent of the lattice site and time step. In our approach, we introduce a site-dependent coin~\cite{Edgard2021}, where the coin’s parameters depend on the lattice site, and a step-dependent coin, where the coin parameters change with time steps~\cite{Mosseri2004,Soriano2006,Schreiber2011,Rigolin2013}. These modifications allow for richer dynamics and greater control over the evolution of the walker's state. Our findings not only challenge the conventional understanding of the requirements for Parrondo's paradox in quantum walks but also open up new avenues for exploring quantum strategies and their applications in quantum transport. The site- and time-dependent coins used in our formalism have been motivated by past experimental setups where such settings were implemented. The ability to induce such paradoxical behavior without additional complexity enhances the feasibility of practical implementations in quantum technologies.

The paper is organized as follows: In Section~\ref{sec:dtqw}, we provide a brief overview of discrete-time quantum walks and the theoretical framework underlying Parrondo's paradox. Section~\ref{sec:parrondo} details our methodology, including the construction of site- and time-dependent coins. In Section~\ref{sec:site-dependent} and Sec~\ref{sec:time-dependent}, we present our results, showcasing the emergence of Parrondo's paradox in the quantum walk model. Finally, we discuss the implications of our findings and potential future research directions and conclude in Section~\ref{sec:conclusion}.

\section{1D DTQW} 
\label{sec:dtqw}
A discrete-time quantum walk is the quantum analog of a classical random walk, characterized by the movement of a quantum particle (the walker) on a lattice, governed by a quantum coin~\cite{Aharonov1993,Ambainis2001,Kempe2003,Nayak2000}. Due to the quantum nature of the walker and the coin, the position state of the walker is a superposition of multiple lattice sites. This provides a quadratically fast spread of the walker across the lattice compared to its classical counterpart \cite{Ambainis2001}. As opposed to classical random walks, quantum walks are governed by quantum superpositions of amplitudes rather than classical probability distributions. The quantum walk setup has been physically realized using trapped atoms and trapped ions~\cite{Milburn2002, Schmitz2009, Roos2010, Karski2009}, waveguide arrays~\cite{Regensburger2012},
photonic setups~\cite{Schreiber2010, Schreiber2011, Regensburger2012, Broome2010, Zhang2007, Sephton2019}, nuclear magnetic resonance~\cite{Du2003, Laflamme2005}, Bose-Einstein condensates~\cite{Sandro2017} orbital angular momentum states of light~\cite{Sephton2019} and photons~\cite{Galton1998}.

One-dimensional (1D) DTQW is defined as a particle hopping over a 1D lattice with an internal degree of freedom, spin, for example, which is equivalent to the coin in a classical random walk. The unit step of a quantum walk consists of a conditional shift operator $\mathcal{T}$ and a coin flip operator $\mathcal{C}(\theta)$ for a real parameter $\theta$. We can represent these operators in the position basis of the lattice $\{\ket{n}\} \in \mathcal{H}_{L}$ and spin basis $\{\ket{\uparrow}, \ket{\downarrow} \} \in \mathcal{H}_C$ of the coin. In this basis, the conditional translation operator $\mathcal{T}$ translates the quantum walker with spin $\ket{\uparrow}$ and $\ket{\downarrow}$ to one lattice site forward and one lattice site backward, respectively. The composite Hilbert space of the coin-lattice system can be written as $\mathcal{H} = \mathcal{H}_C \otimes \mathcal{H}_L$. The conditional shift (or translation) and coin operators read as follows
\begin{equation}
    \mathcal{T} = \sum_x \dyad{\uparrow} \otimes \dyad{x+1}{x}  + \dyad{\downarrow} \otimes \dyad{x-1}{x}, 
\end{equation}
and
\begin{equation}
    \mathcal{C}(\theta) = e^{-i \theta \sigma_y/2} \otimes \mathds{1}_{N},
\end{equation}
where $-2\pi\le\theta<2\pi$ is a real parameter and $\sigma_y$ is the Pauli matrix along the $y$-axis. Here, $\mathds{1}_N$ represents the identity operation on the lattice with $N$ sites. The unitary operator, which governs the time evolution of the walker for a unit time step, reads
\begin{equation}
	U(\theta) = \mathcal{T} \mathcal{C}(\theta).
	\label{eq:unitaryevolution}
\end{equation}

\begin{figure*}
	\centering
	\subfigure[]{\includegraphics[width=0.23\textwidth]{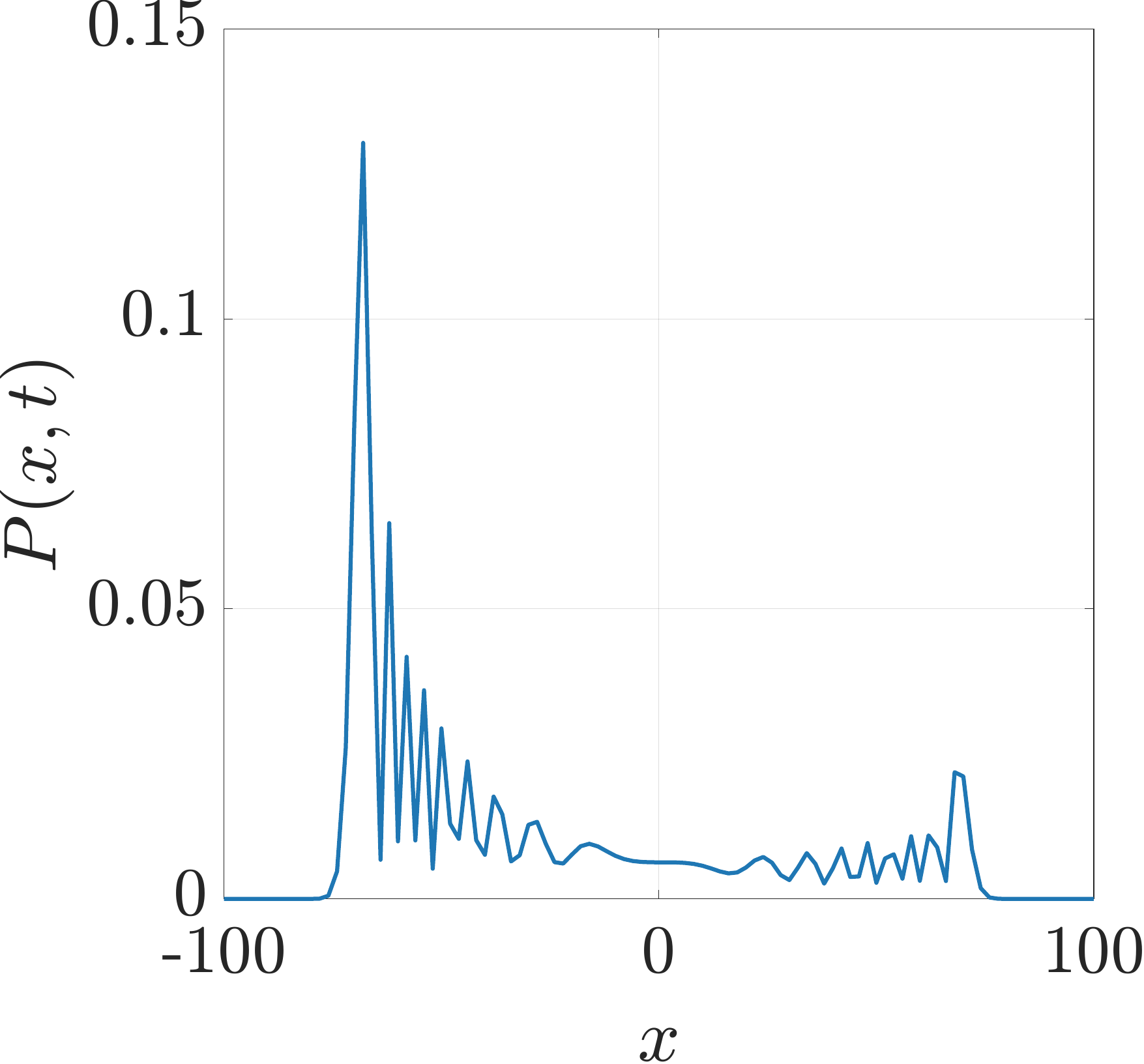}
		\label{fig:1DEvolutiondown}}
	\subfigure[]{\includegraphics[width=0.23\textwidth]{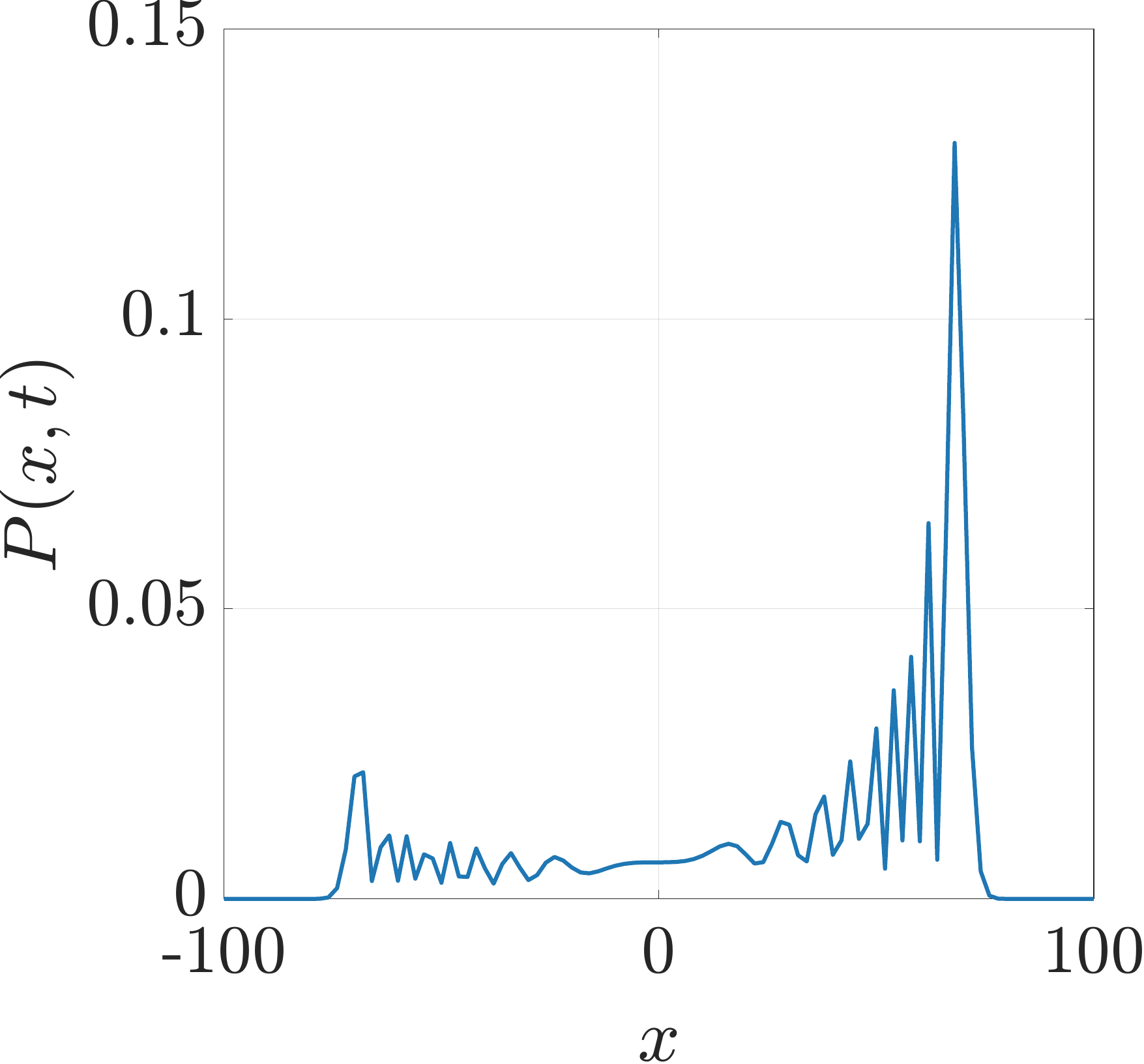}
		\label{fig:1DEvolutionup}}
	\subfigure[]{\includegraphics[width=0.23\textwidth]{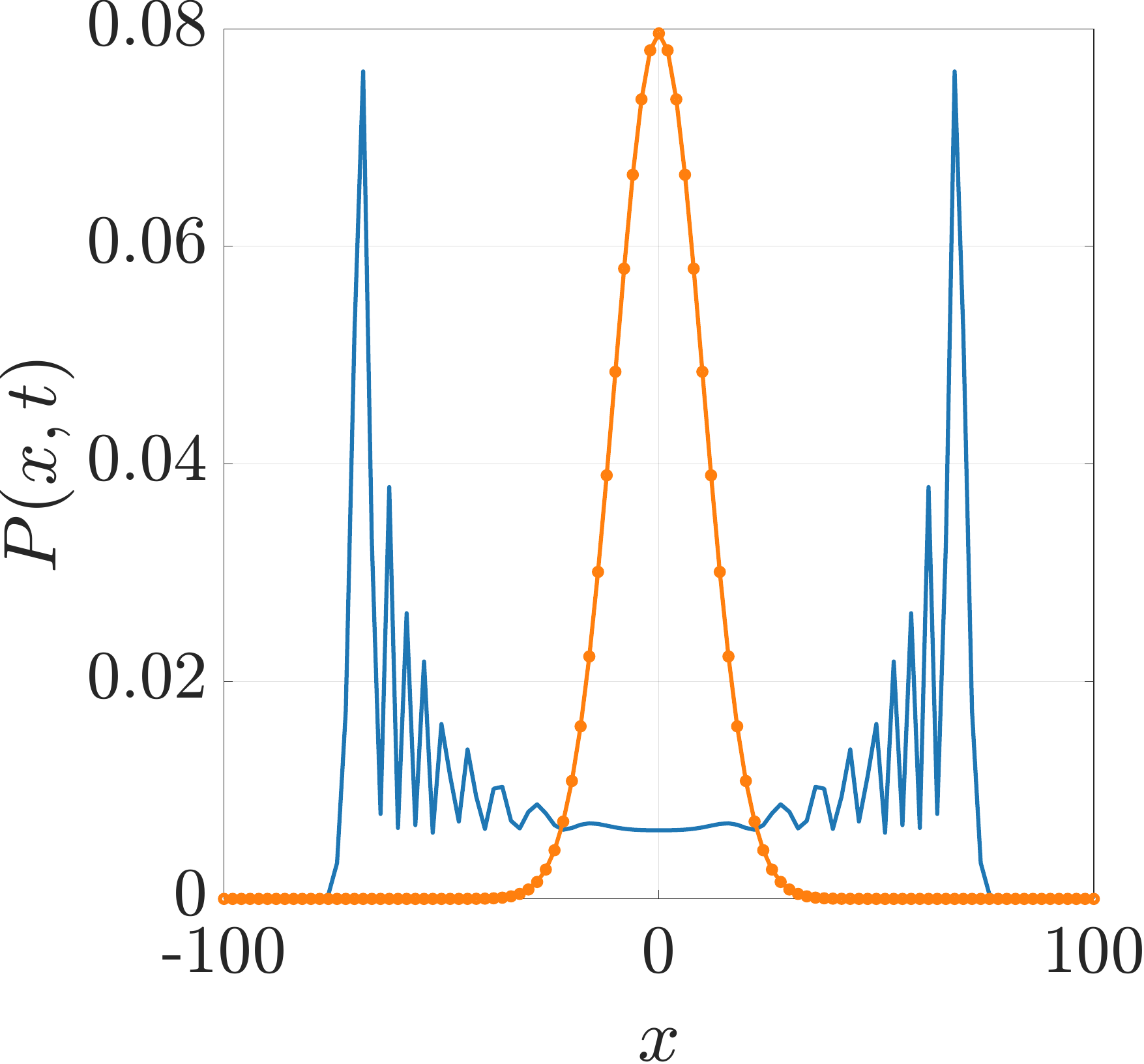}
		\label{fig:dtqw}}	
	\subfigure[]{\includegraphics[width=0.23\textwidth]{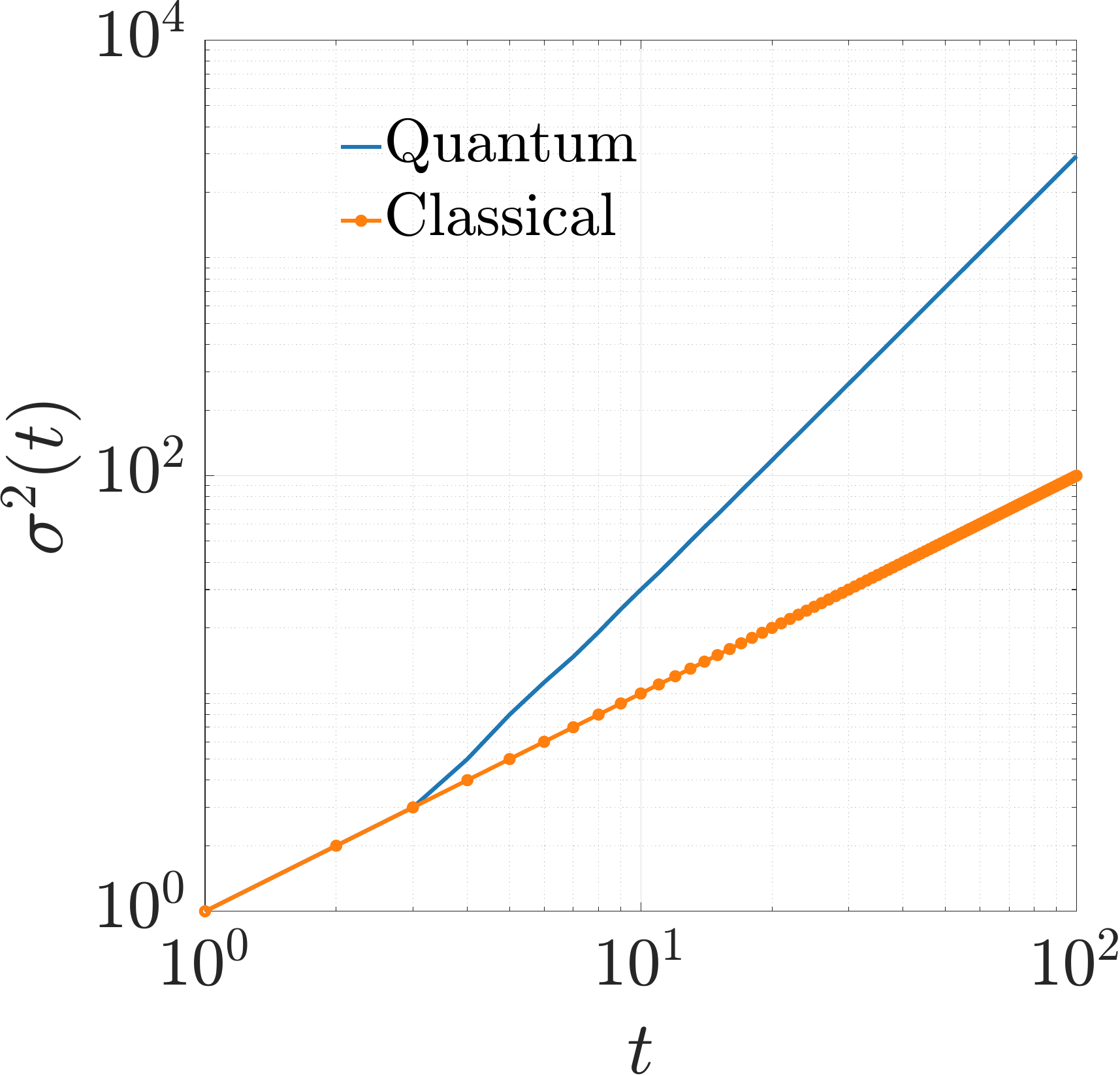}
		\label{fig:variance}}
	\caption{Probability distribution of a walker for a $1$D DTQW with $N = 201$ lattice sites after $t = 100$ time steps for different initial states. \subref{fig:1DEvolutiondown} $ \ket{\psi(0)} = \ket{\downarrow}_C \otimes \ket{0}_L $, \subref{fig:1DEvolutionup} $ \ket{\psi(0)} = \ket{\uparrow}_C \otimes \ket{0}_L $, \subref{fig:dtqw} $ \ket{\psi(0)} = (\ket{\uparrow}_C + i \ket{\downarrow}_C)/\sqrt{2} \otimes \ket{0}_L$ in contrast to classical walk (red). The rotation angle in the coin flip operator $\mathcal{C}(\theta)$ is chosen to be $\theta = \pi/2$ for all the plots. \subref{fig:variance} The variance, $\sigma^2(t)$ for quantum and classical walks as a function of time steps $t$ on a logarithmic scale. In the subplots (a), (b), and (c), only the points with non-zero probability are plotted for better presentation.} 
	\label{fig:ProbDis}
\end{figure*}

The state of the walker after $t$ time steps starting from $ \ket{\psi(0)} $ is denoted as
\begin{equation}
	\begin{aligned}
		\ket{\psi(t)} &= (U(\theta))^t \ket{\psi(0)} \\
		&= \sum_x \psi_{\uparrow, x} (t) \ket{\uparrow} \otimes \ket{x} + \psi_{\downarrow, x} (t) \ket{\downarrow} \otimes \ket{x},
	\end{aligned}  
\end{equation}
and the probability of finding the walker at the $x$th lattice site after $t$ time steps is given by
\begin{align}
	P(x,t) &= \abs{\bra{\uparrow} \otimes \ip{x}{\psi(t)}}^2 + \abs{\bra{\downarrow} \otimes \ip{x}{\psi(t)}}^2 \nonumber \\ 			
	&= \abs{\psi_{\uparrow, x} (t)}^2 + \abs{\psi_{\downarrow, x} (t)}^2.
	\label{eq:probdist}
\end{align}
where $\psi_{\uparrow, x} (t)$ and $\psi_{\downarrow, x} (t)$ are the amplitudes of the lattice state at time $t$ corresponding to the two coin states $\ket{\uparrow}$ and $\ket{\downarrow}$ respectively. For a given probability distribution $P(x,t)$ of the quantum walker, the average position of the wavefunction, or equivalently, the expectation value of the position operator, $\hat{X}$ defined as 
\begin{equation}
\label{eq:expectation}
	\expval*{\hat{X}(t)} = \sum_x x P(x, t).
\end{equation}
This quantity will be of importance when we discuss the winning criterion in quantum walk games in upcoming sections. In Figures~\ref{fig:1DEvolutiondown},~\ref{fig:1DEvolutionup}, and~\ref{fig:dtqw}, we present a detailed analysis of the probability distribution of the walker, as described by Eq.~\eqref{eq:probdist}, evolving over 100 time steps across various initial states. Figure~\ref{fig:dtqw} specifically compares the probability distributions observed in classical and quantum walks, highlighting their distinct behaviors. Quantum walks clearly show the dynamic evolution of probability amplitudes over time, resulting in notable constructive and destructive interference patterns. Constructive interference is especially prominent at the edges of the lattice, while significant destructive interference occurs at the origin. In contrast, classical random walks follow a predictable Gaussian distribution pattern without any interference effects.

Figure~\ref{fig:variance} illustrates the variance of the probability distribution, denoted as $\sigma^2 (t)$, within a 1D DTQW with a specific initial state. The variance increases quadratically with the number of time steps $t$, following the relationship $\sigma^2(t) \propto t^2$. This growth rate is twice as fast as that observed in classical random walks. Note, in figure~\ref{fig:variance}, we use a logarithmic scale to show the variance trends in both classical and quantum walks. Consequently, the quantum walk and classical walk are often referred to as ballistic and diffusive processes, respectively, highlighting their inherent dynamics and variance characteristics.

\section{Parrondo's Paradox	}
\label{sec:parrondo}
Parrondo's paradox is an intriguing phenomenon where two losing strategies or games, when alternated or combined, result in a winning outcome. Extending this paradox to the realm of quantum mechanics, specifically in the setting of 1D DTQW, involves defining analogous strategies using quantum coins as players and determining the conditions under which a winning outcome is achieved.

In a quantum walk with discrete-time dynamics, the progression of the walker is determined through the application of a coin operator followed by a shift operator. The walker's state comprises a superposition of positional states influenced by the coin operator, which governs the direction of movement. To introduce Parrondo's paradox within the context of quantum walks, it is necessary to establish multiple quantum games involving a minimum of two distinct games. Conventionally, quantum games within quantum walk contexts are characterized by the selection of a specific coin operator. By utilizing two coin operators, denoted as \Coin{A} and \Coin{B}, we create two distinct quantum games. The choice of a particular coin operator is analogous to picking a strategy for engaging in the game and advancing the quantum walk. The two games individually lead to losing outcomes when applied repeatedly. Before moving further, we first define the winning criterion in the games. 

To quantify the winning or losing nature of these games or strategies, we use the expectation value of the position operator defined in Eq.~\ref{eq:expectation}. A positive drift in the expectation value over time indicates a winning strategy, whereas a negative drift signifies a losing one. In the context of Parrondo's paradox, even though $\expval*{\hat{X}(t)}$ decreases when either $\mathcal{C}_A$ or $\mathcal{C}_B$ is applied individually (signifying losing outcomes), the combined sequence of these coin operators might results in an increasing $\expval*{\hat{X}(t)}$, indicating a net winning outcome. Now, we define the two quantum games as follows:
\begin{enumerate}
    \item \textbf{Game A (Coin $\mathcal{C}_A$):} In this game, the coin operator $\mathcal{C}_A$ is applied at each time step of the quantum walk. The coin operator $\mathcal{C}_A$ is a unitary matrix that, when used consistently, results in a quantum walk that exhibits a drift towards negative $x$-direction, leading to a net losing outcome.
    
    \item \textbf{Game B (Coin $\mathcal{C}_B$):} This game uses a different coin operator $\mathcal{C}_B$ at each time step. Similar to $\mathcal{C}_A$, the coin operator $\mathcal{C}_B$ also results in a biased quantum walk in a negative $x$ direction, leading to another losing outcome.
\end{enumerate} 

Having defined the terminology for individual games, we now detailed our notation for combined strategies. We define the combined game or strategy as follows: 
\begin{equation}
    \label{eq:combineddef}
    \mathcal{C}_{AB}^{(m,n)}  \equiv \mathcal{C}_A^m \mathcal{C}_B^n
\end{equation}
where $m, n = 0, 1, 2, \dots n$. This translates to the application of \Coin{A} $m$-times, followed by the application of \Coin{B} $n$-times in a single time step of a quantum walk. The different pairs of $(m,n)$ correspond to distinct quantum games or strategies.   

The paradox highlights how quantum strategies that are individually disadvantageous can combine to produce results that are not intuitive, broadening our understanding of quantum dynamics and offering potential applications in quantum computation and information processing. In the next section, we detail our construction of deterministic combined sequences using the individual games and leave the probabilistic construction for Appendix~\ref{appen:probparr}.

\begin{figure*}
	\centering
	\subfigure[]{\includegraphics[height=0.21\textwidth]{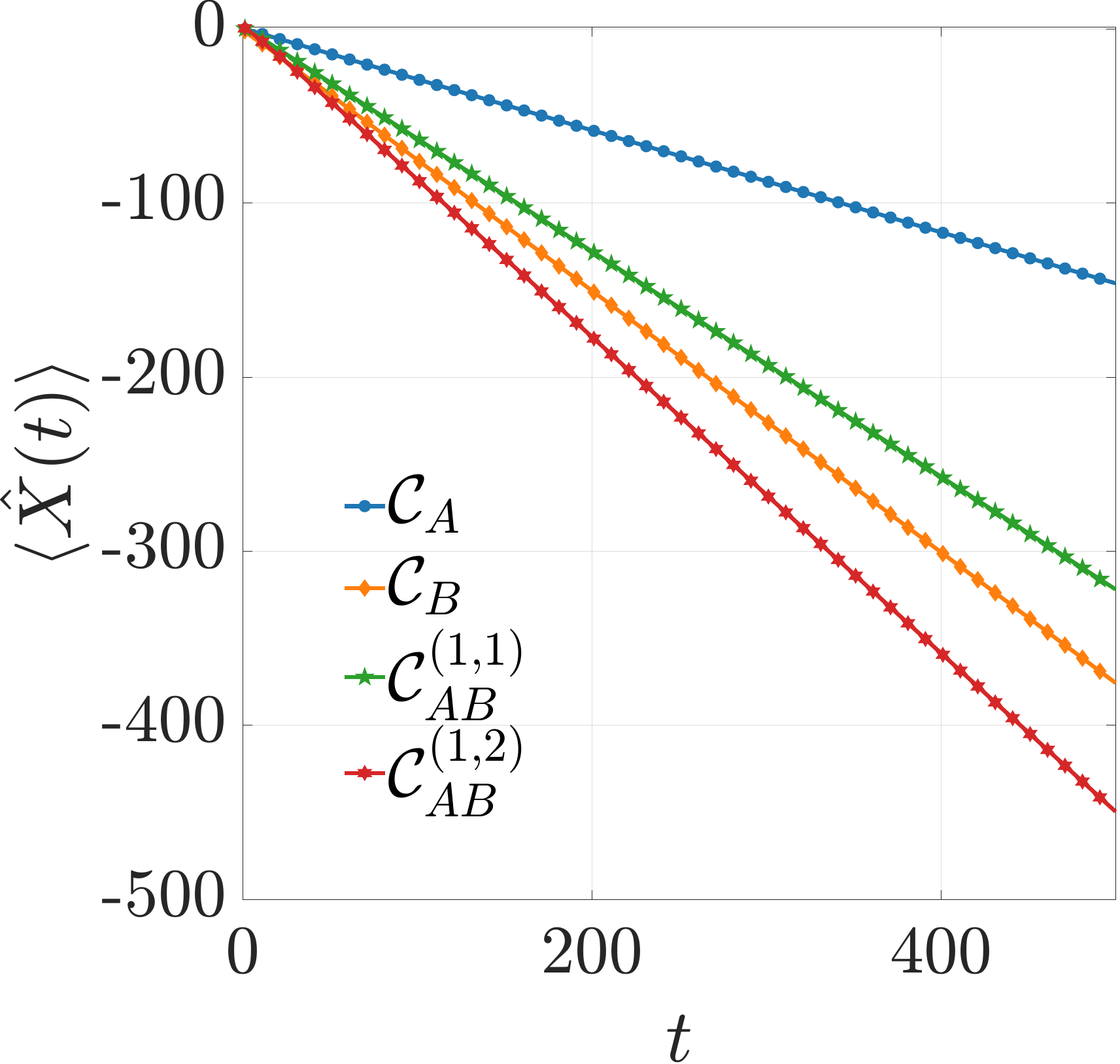}}
	\subfigure[]{\includegraphics[height=0.21\textwidth]{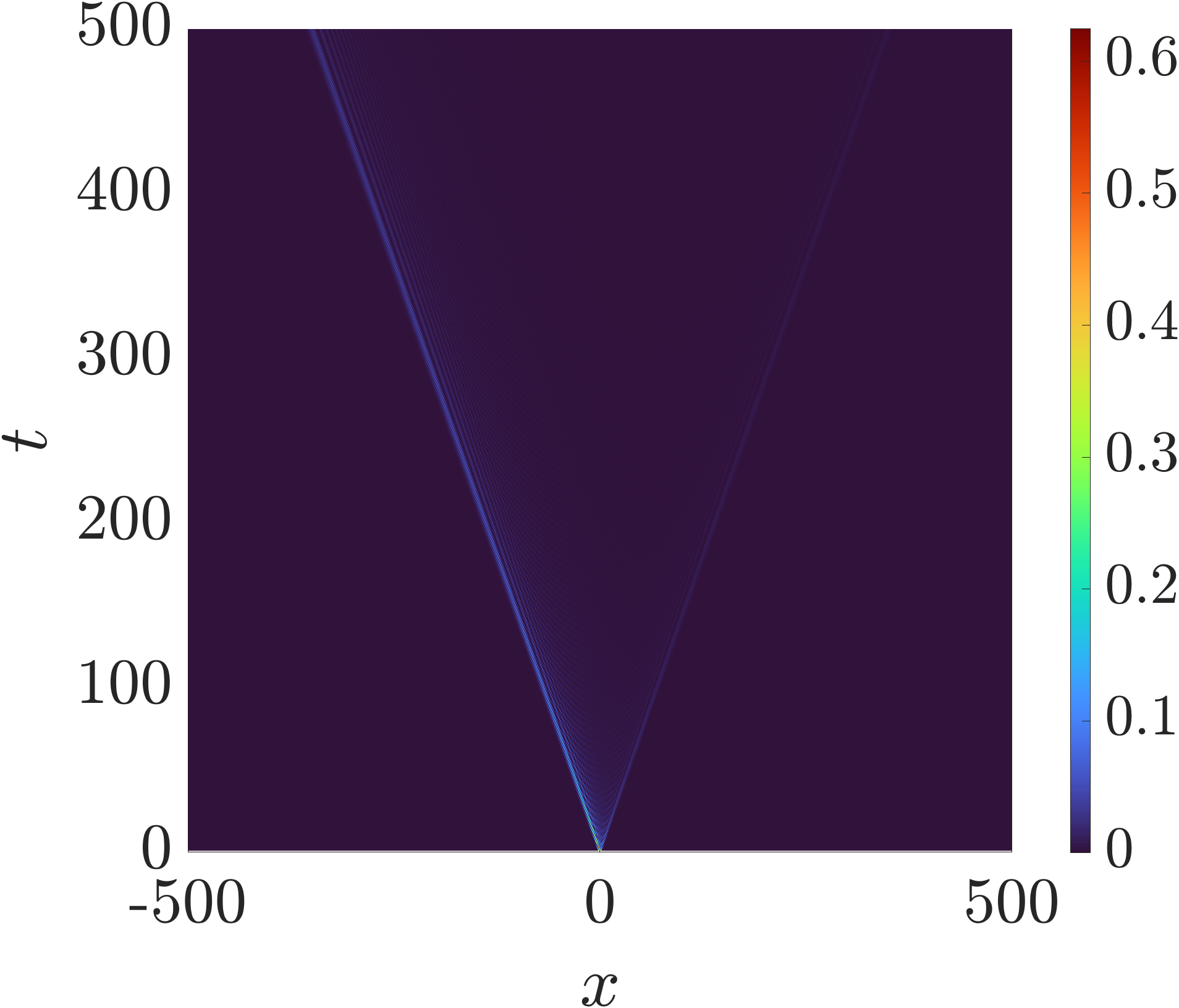}}
	\subfigure[]{\includegraphics[height=0.21\textwidth]{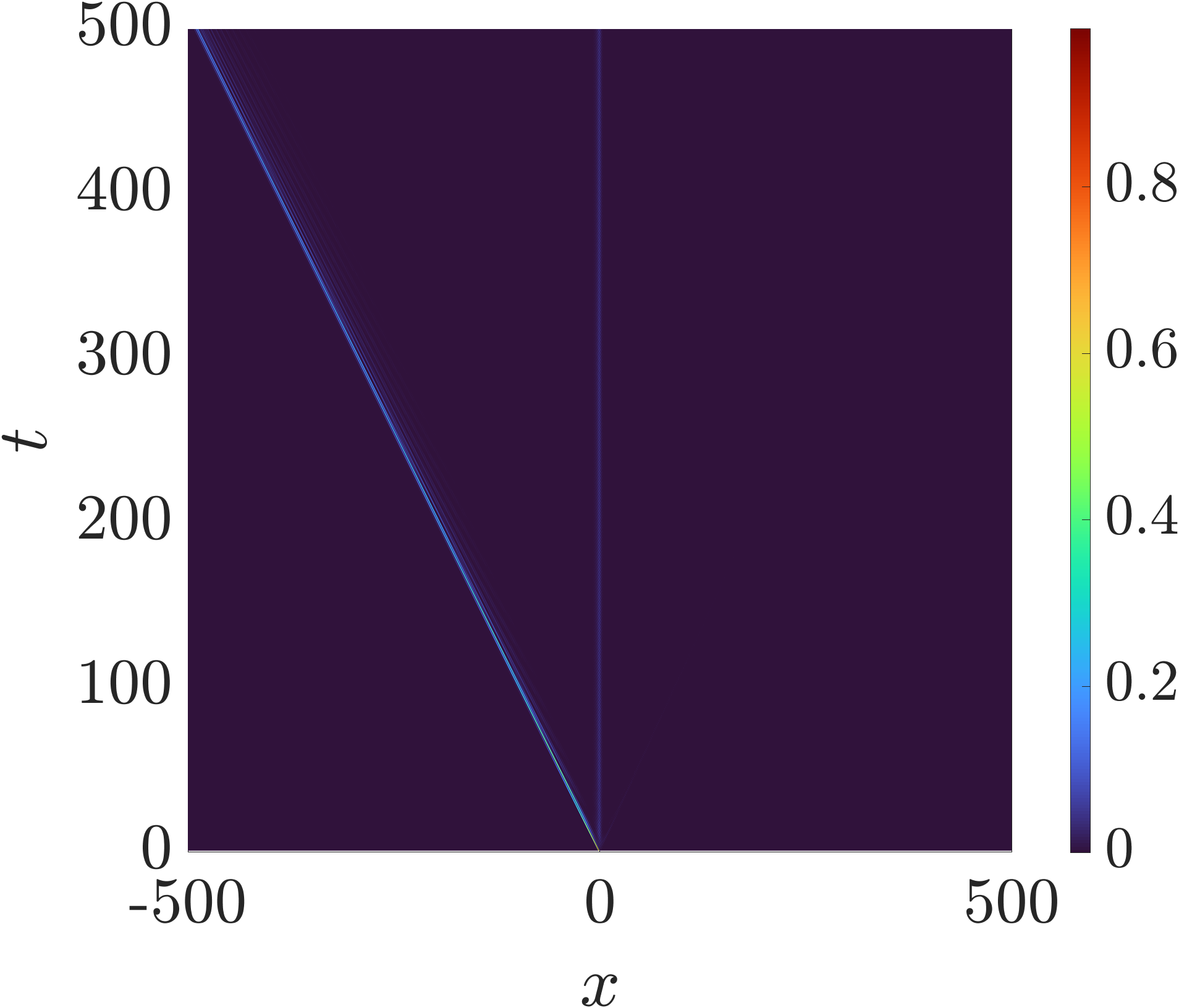}}
	\subfigure[]{\includegraphics[height=0.21\textwidth]{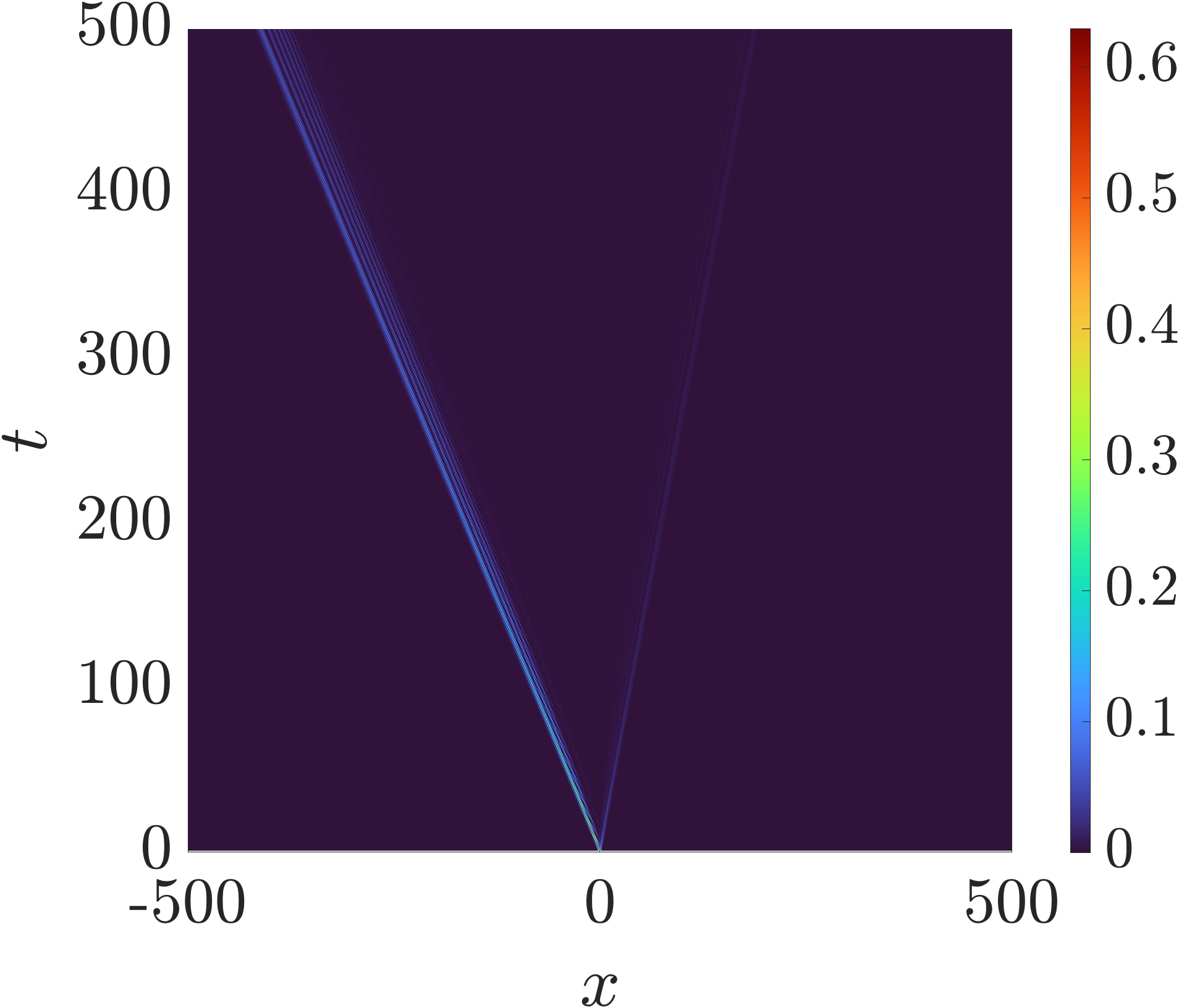}}
	
	\subfigure[]{\includegraphics[height=0.21\textwidth]{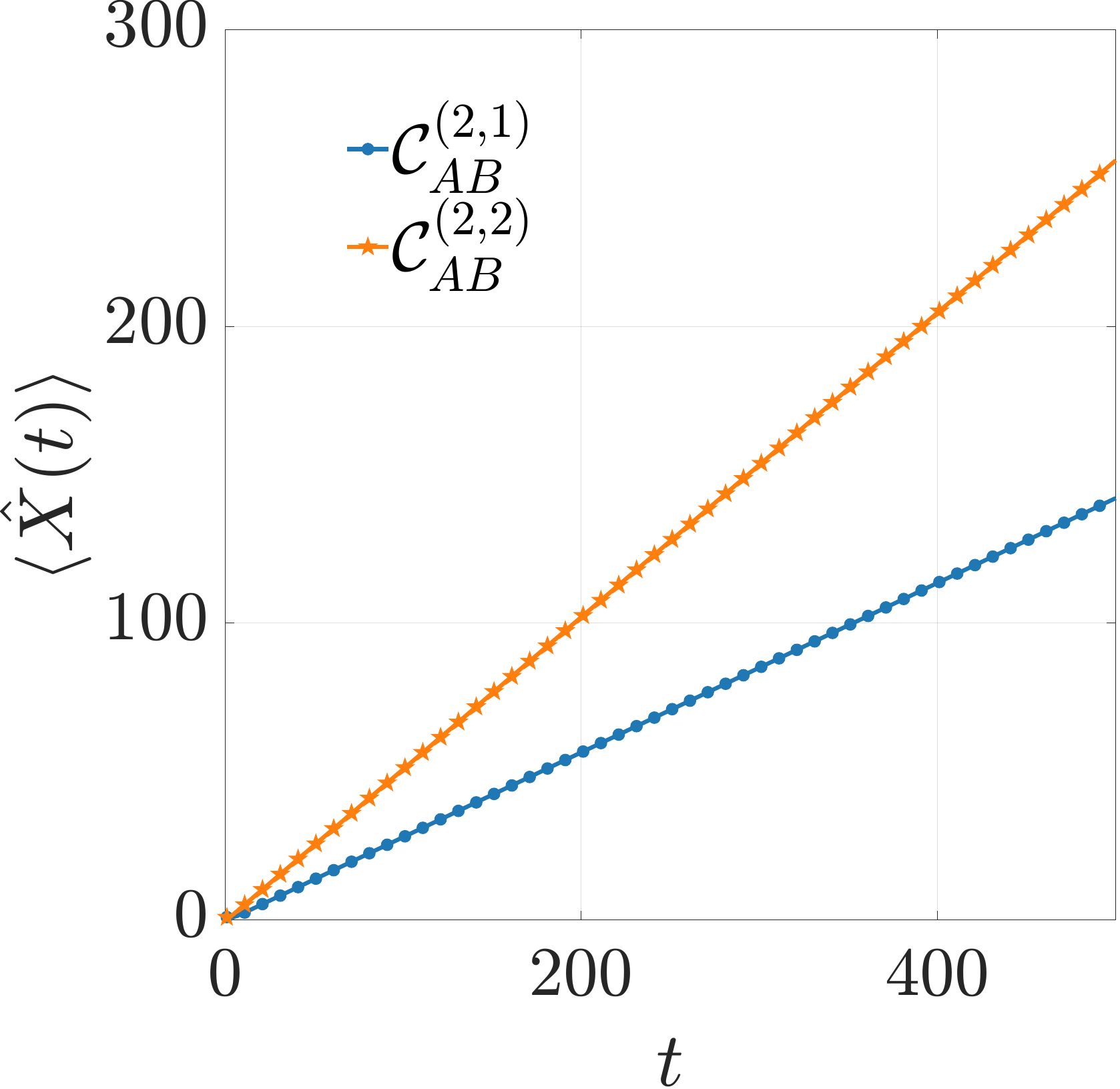}}
	\subfigure[]{\includegraphics[height=0.21\textwidth]{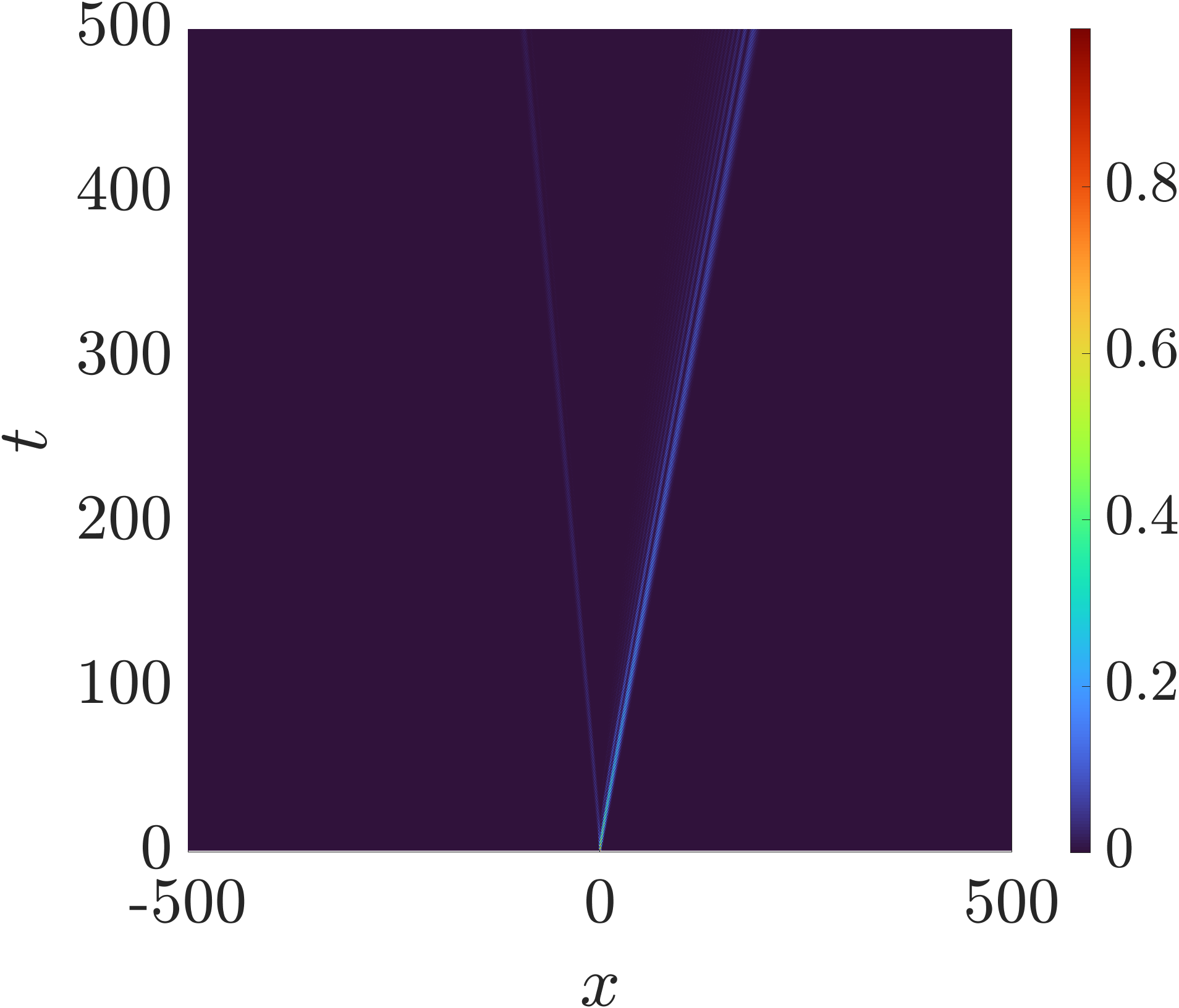}}
	\subfigure[]{\includegraphics[height=0.21\textwidth]{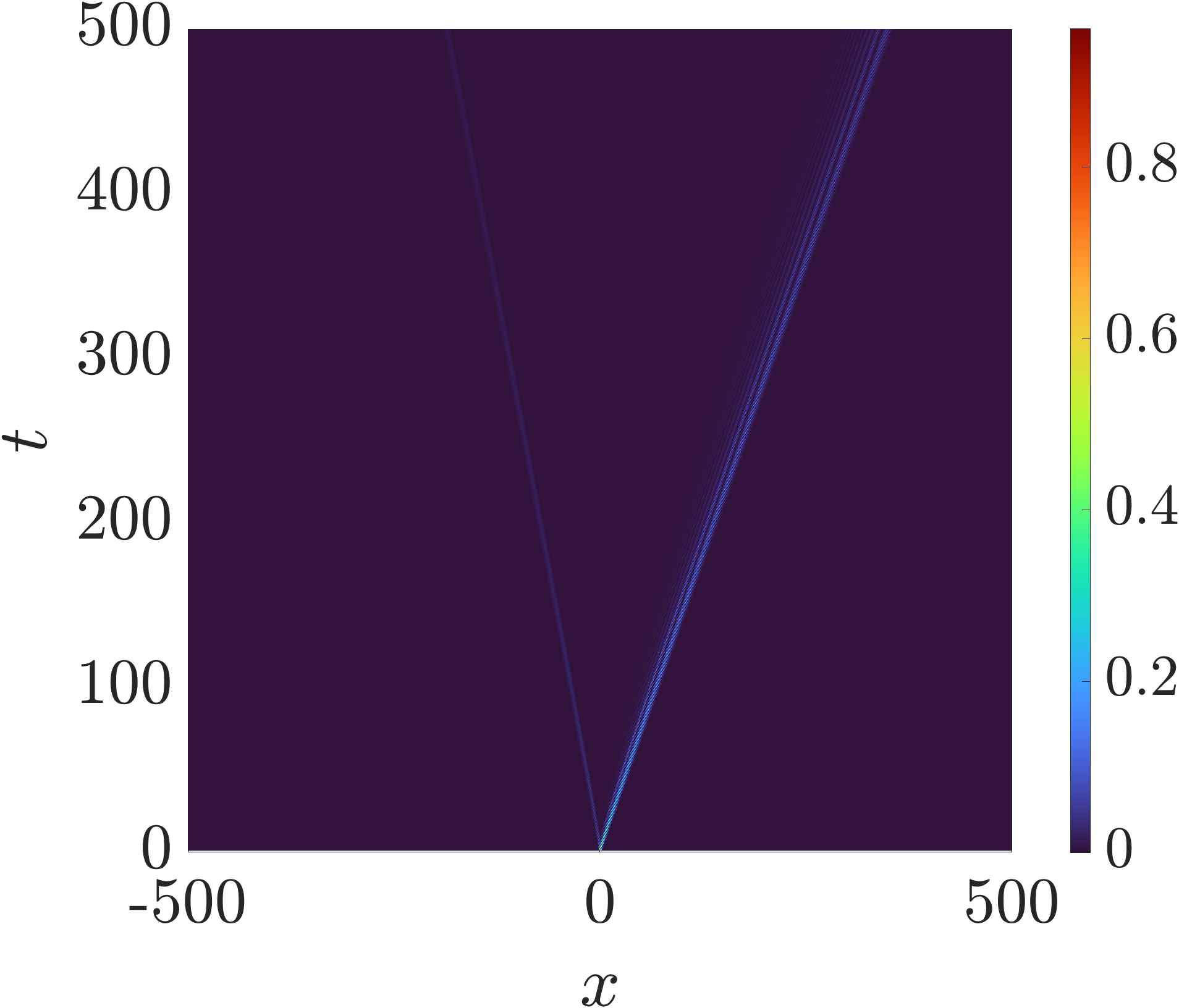}}
        \caption{The expectation value and the probability distribution of the walker as a function of time for different sequences of coins. The two individual coins or quantum games $\mathcal{C}_A$ and $\mathcal{C}_B$ are characterized by $\theta_A = \pi/2$ and $(\theta_{B-}, \theta_{B+}) = (-\pi/8, \pi/4)$ respectively. (a) the expectation value of the position of the quantum walker for individual coins $\mathcal{C}_A$, $\mathcal{C}_B$ and a trivial combination of two coins $\mathcal{C}_{AB}^{(1,1)}$ = \Coin{A}\Coin{B}, and $\mathcal{C}_{AB}^{(1,2)}$ = $\mathcal{C}_A\mathcal{C}_B^2$ (b), (c), and (d) is the probability distribution of the quantum walkers with coins $\mathcal{C}_A$, $\mathcal{C}_B$ and $\mathcal{C}_{AB}^{(1,1)}$ respectively. Note: We are not showing the time evolution of the probability for $\mathcal{C}_{AB}^{(1,2)}$ in order to avoid clutter. In all the cases, the probability current of the quantum walker is biased towards negative $x$ at all times, which results in a losing strategy for these cases. Now we consider the non-trivial (that leads to a portability current in the opposite direction and hence leads to a winning strategy) sequence of coins:  $\mathcal{C}_{AB}^{(2,1)}$ = $\mathcal{C}_A^2$\Coin{B} and $\mathcal{C}_{AB}^{(2,2)}$ = $\mathcal{C}_A^2 \mathcal{C}_B^2$. (e) the expectation value of the position of the position operator where we observe a counter-intuitive shift in probability distribution. (f) and (g) represent the probability distribution of quantum walker with time for coin sequences $\mathcal{C}_{AB}^{(2,1)}$ and $\mathcal{C}_{AB}^{(2,2)}$ respectively. The system size is taken to be $N = 1001$. The initial state of the composite system is considered to be $\ket{\Psi(0)} = \ket{\psi}_C \otimes \ket{\psi}_L = \ket{\downarrow} \otimes \ket{0}$.}
	\label{fig:site-dependent}
\end{figure*}

\section{Site-Dependent Coin}
\label{sec:site-dependent}
In this section, we detail our construction to realize Parrondo's paradox using the site-dependent coin in discrete-time quantum walks. Let us consider two distinct quantum coins \Coin{A} and \Coin{B}, each corresponding to different quantum games. When applied independently, these coins result in losing strategies, meaning that the expectation value of the position operator leads to a net negative outcome. 
\begin{enumerate}
    \item \textbf{Game A (Coin \Coin{A}):} \\
    \noindent The coin operator \Coin{A} represents the first losing strategy. It is defined as a simple SU(2) rotation along the $y$-axis, but when applied over several steps, it fails to produce a distribution for the walker with the positive expectation value of the position operator. It reads
    \begin{equation}
        \label{eq:coinA}
        \mathcal{C}_A = e^{-i \theta_A \sigma_y/2} \otimes \mathds{1}_{N}
    \end{equation}
    where $\theta \in [-2 \pi, 2 \pi]$ is chosen such that the resulting quantum walk exhibits a losing strategy.
    \item \textbf{Game B (Coin \Coin{B}):} \\
    The coin operator \Coin{B} represents the second losing strategy. Again, we define this as a simple SU(2) rotation along the $y$-axis, but with a rotation angle dependent on the lattice sites. It reads
    \begin{equation}
        \label{eq:coinB}
        \mathcal{C}_B = \sum_x e^{-i \theta_B(\theta_{B_+},\theta_{B_-};x) \sigma_y/2} \otimes \dyad{x}
    \end{equation}
    where 
    \begin{equation}
    \begin{aligned}
     &\theta_B(\theta_{B_+},\theta_{B_-};x) \\
     &= \frac{\theta_{B_+}\left(1+\tanh(x)\right)+\theta_{B_-}\left(1-\tanh(x)\right)}{2}.
    \end{aligned}
    \end{equation}
    Here, as before, $\theta_B{\pm}$ are chosen such that the resulting quantum walk exhibits a losing outcome.
    Such site-dependent rotations have already been employed experimentally in photonic quantum walk setup~\cite{Kitagawa2012}. 
\end{enumerate}
When the walker undergoes a quantum walk with only coin \Coin{A} applied at each step, the state evolution is governed by the following time evolution operator
\begin{equation}
	U_A = \mathcal{T}\mathcal{C}_A.
\end{equation}
With a particular choice of $\theta_A$, over time, the probability distribution of the walker's position under $U_A$ does not result in positive drift and hence corresponds to a losing situation.

Similarly, for coin \Coin{B}, the state evolution is governed by the following time evolution operator
\begin{equation}
	U_B = \mathcal{T}\mathcal{C}_B
\end{equation}
which also leads to negative drift. 

We now consider a numerical simulation of the individual quantum walk. The initial state of lattice and coin combined system, $\ket{\Psi(0)}$ is localized at the origin and it reads 
\begin{equation}
	\ket{\Psi(0)} = \ket{\downarrow} \otimes \ket{0}.
\end{equation} 
Further, the coin parameters for Games A and B are chosen to be $\theta_A = \pi/2$ and $(\theta_{B-}, \theta_{B+}) = (-\pi/8, \pi/4)$ respectively. We extended our analysis to examine the effects of various initial states of the quantum walker in Appendix~\ref{appen:initialstate}, where we show the existence of several regions in the phase space of coin parameters where the manifestation of Parrondo's effect can be observed. 

After applying the individual unitary operations defined at the beginning of this section for a sufficient number of steps, the resulting probability distribution $P(x,t)$ and the expectation value of the position, $\expval*{\hat{X}(t)}$ are plotted in Fig.~\ref{fig:site-dependent}. In the top row of Fig.~\ref{fig:site-dependent}, we plot the expectation value of the position, $\expval*{\hat{X}(t)}$, and the probability distribution as a function of time. As it is evident from the plots, there is a higher probability of finding the walker at the negative positions compared to the initial state. We observe that the expectation value of the position operator keeps on decreasing with time $t$, and it is impossible for us to win this game independently with the individual coins $\mathcal{C}_A$ and $\mathcal{C}_B$. We discuss the probability distribution for individual and combined strategies for the first few steps in Appendix~\ref{appen:dynamics}.

\subsection*{Winning Strategy}
The central idea of Parrondo's paradox is that two individually losing strategies can be combined in a specific sequence to produce a winning outcome. For our purpose, we define the following deterministic sequence, which is combined using the individual games with coins \Coin{A} and \Coin{B} as 
\begin{equation}
	\label{eq:combined1}
	\mathcal{C}_{AB}^{(2,1)} = \mathcal{C}_A^2 \mathcal{C}_B.
\end{equation}
where \Coin{A} and \Coin{B} are the same coin operators defined in Eq.~\ref{eq:coinA} and~\ref{eq:coinB}. This notation concisely captures the order of operations applied in each step, ensuring clarity and avoiding clutter. The combined evolution operator for this sequence can be expressed as:
\begin{equation}
    \label{eq:combinedevolution}
    U_{AB} \equiv \mathcal{T} \mathcal{C}_{AB}^{(2,1)}
\end{equation}
where each time step involves using the coin \Coin{A} twice followed by \Coin{B} once. The above sequence introduces constructive interference patterns that lead to a net positive bias in the walker's position. This constructive interference is not present when either \Coin{A} or \Coin{B} is applied independently. The effect of the combined strategy modifies the walker's probability distribution in a way that enhances the probability of finding the walker on the positive side of the origin.

To quantitatively analyze the emergence of a winning strategy, we again examine the probability distribution $P(x, t)$ and $\expval*{\hat{X}(t)}$ of the walker's position $x$ with time $t$. The combined strategy is considered winning if $P(x, t)$ shows a significant bias towards certain positions that are defined as winning states. The results of the simulations are plotted in the bottom row of Fig.~\ref{fig:site-dependent}. The plot shows the peak of the probability distribution towards the right, indicating a higher probability of finding the walker at these positions compared to the initial state. These peaks represent the winning outcome induced by the combined strategy. This is evident from the plot of $\expval*{\hat{X}(t)}$ where we see a linear growth over time and an overall probability current towards the positive $x$-direction.

We further introduce another sequence of coin operators as 
\begin{equation}
	\label{eq:combined2}
	\mathcal{C}_{AB}^{(2,2)} = \mathcal{C}_A^2 \mathcal{C}_B^2
\end{equation}
which also results in the probability current shifting towards the right and hence a winning scenario, as shown in Fig.~\ref{fig:site-dependent}. Hence, the two individual strategies with a probability distribution biased toward the left can together result in a scheme that results in a probability distribution biased toward the right. This is the essence of Parrondo's Paradox.

\begin{figure*}
	\centering
	\subfigure[]{\includegraphics[height=0.20\textwidth]{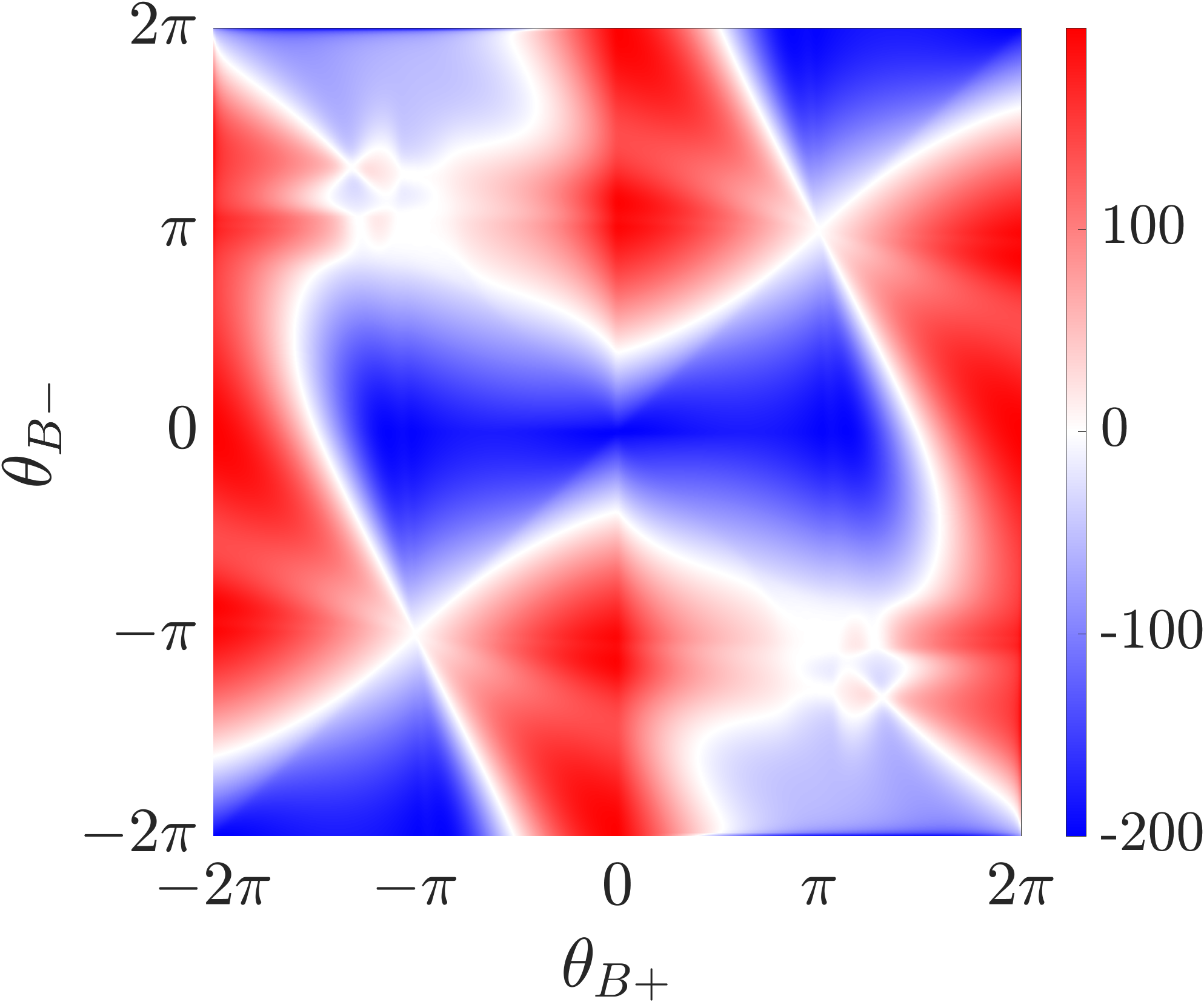}}
	\subfigure[]{\includegraphics[height=0.20\textwidth]{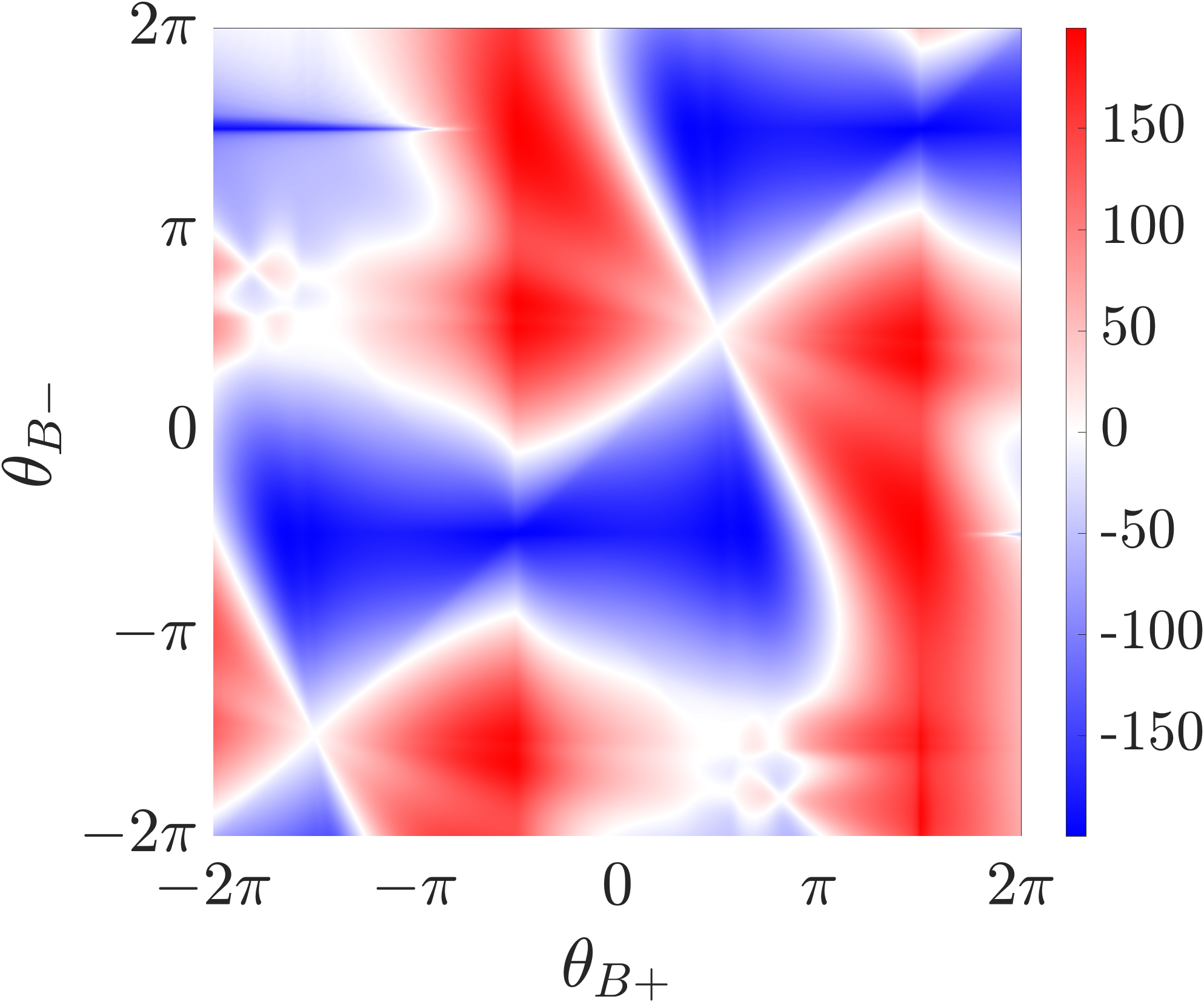}}
	\subfigure[]{\includegraphics[height=0.20\textwidth]{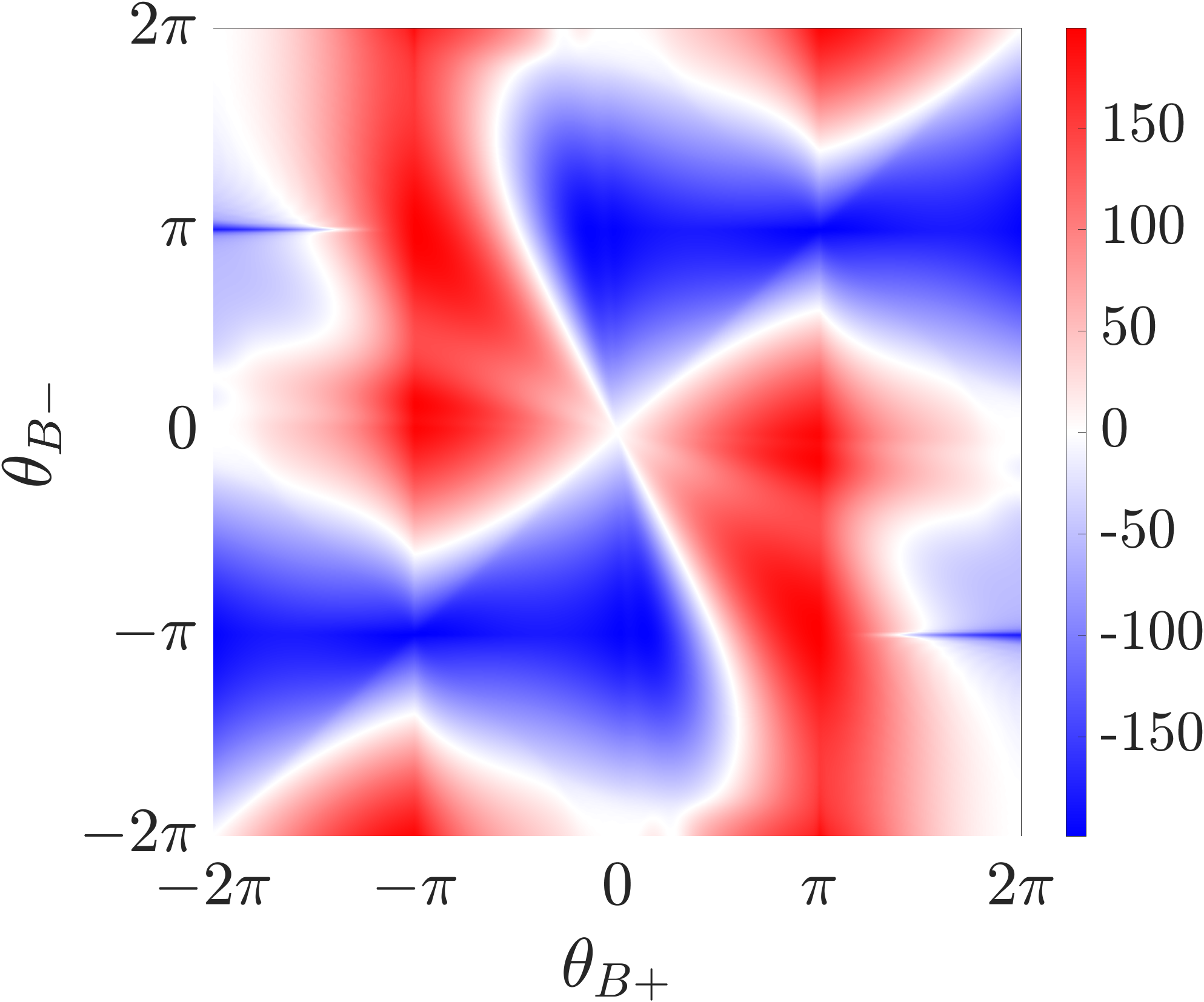}}
	\subfigure[]{\includegraphics[height=0.20\textwidth]{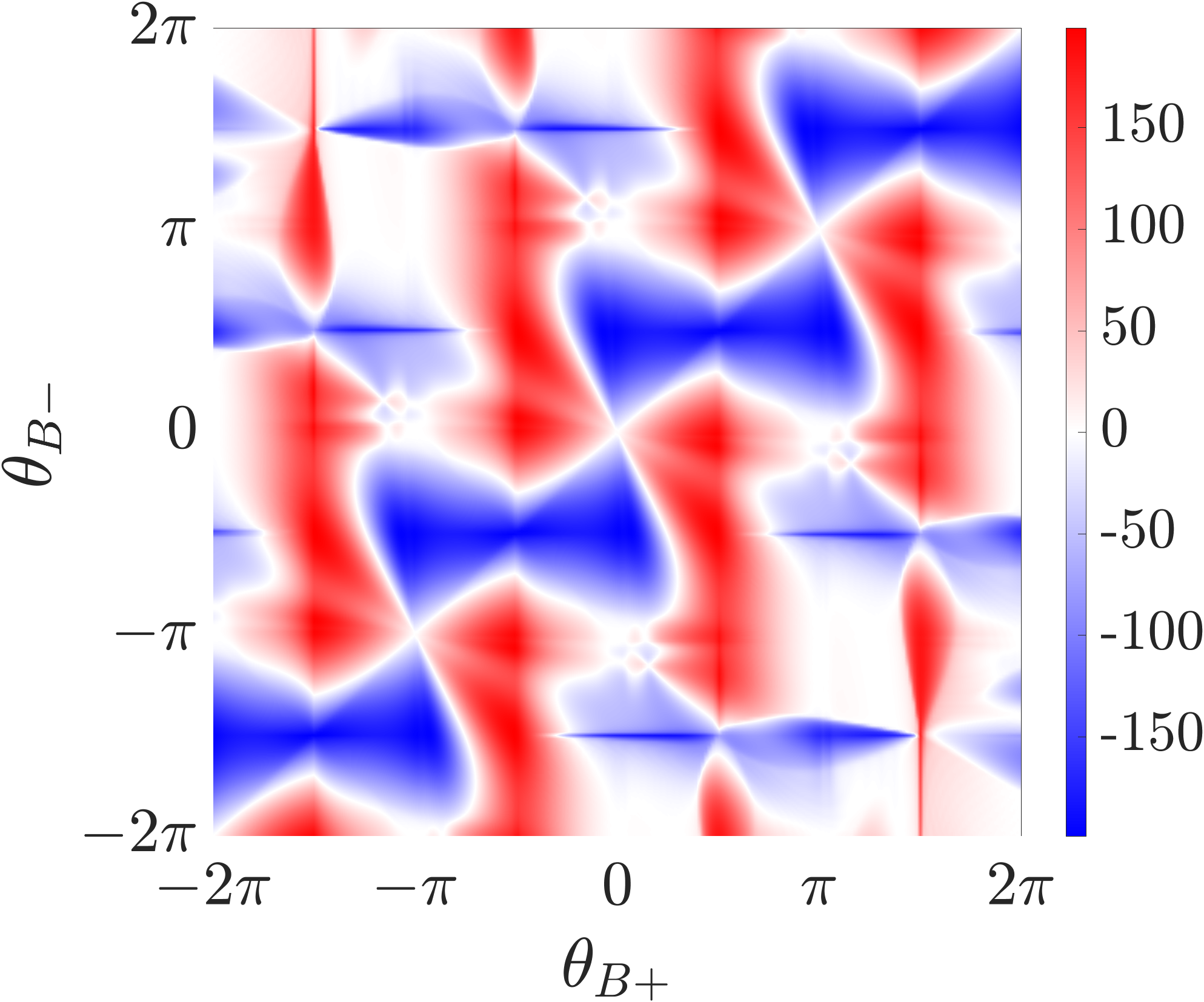}}
	\caption{The expectation value of the position of the walker as a function of the coin parameters $\theta_{B-}$ and $\theta_{B+}$ for the individual strategies (a) \Coin{B}, and the combined sequences (b) $\mathcal{C}_{AB}^{(1,1)} = \mathcal{C}_A \mathcal{C}_B$, (c) $\mathcal{C}_{AB}^{(2,1)} = \mathcal{C}_A^2 \mathcal{C}_B$, and (d) $\mathcal{C}_{AB}^{(2,2)} = \mathcal{C}_A^2 \mathcal{C}_B^2$ for $t = 200$ time steps. The coin parameter, $\theta_A$ for coin \Coin{A}, is kept fixed at $\pi/2$. The \Red{red} and the \Blue{blue} shaded regions correspond to the \Red{winning} and \Blue{losing} situation, respectively. The system size is taken to be $N = 501$. The initial state of the composite is considered to be $\ket{\Psi(0)} = \ket{\psi}_C \otimes \ket{\psi}_L = \ket{\downarrow} \otimes \ket{0}$. }
	\label{fig:thetapm}
\end{figure*}

\begin{figure*}
	\centering
	\subfigure[]{\includegraphics[height=0.20\textwidth]{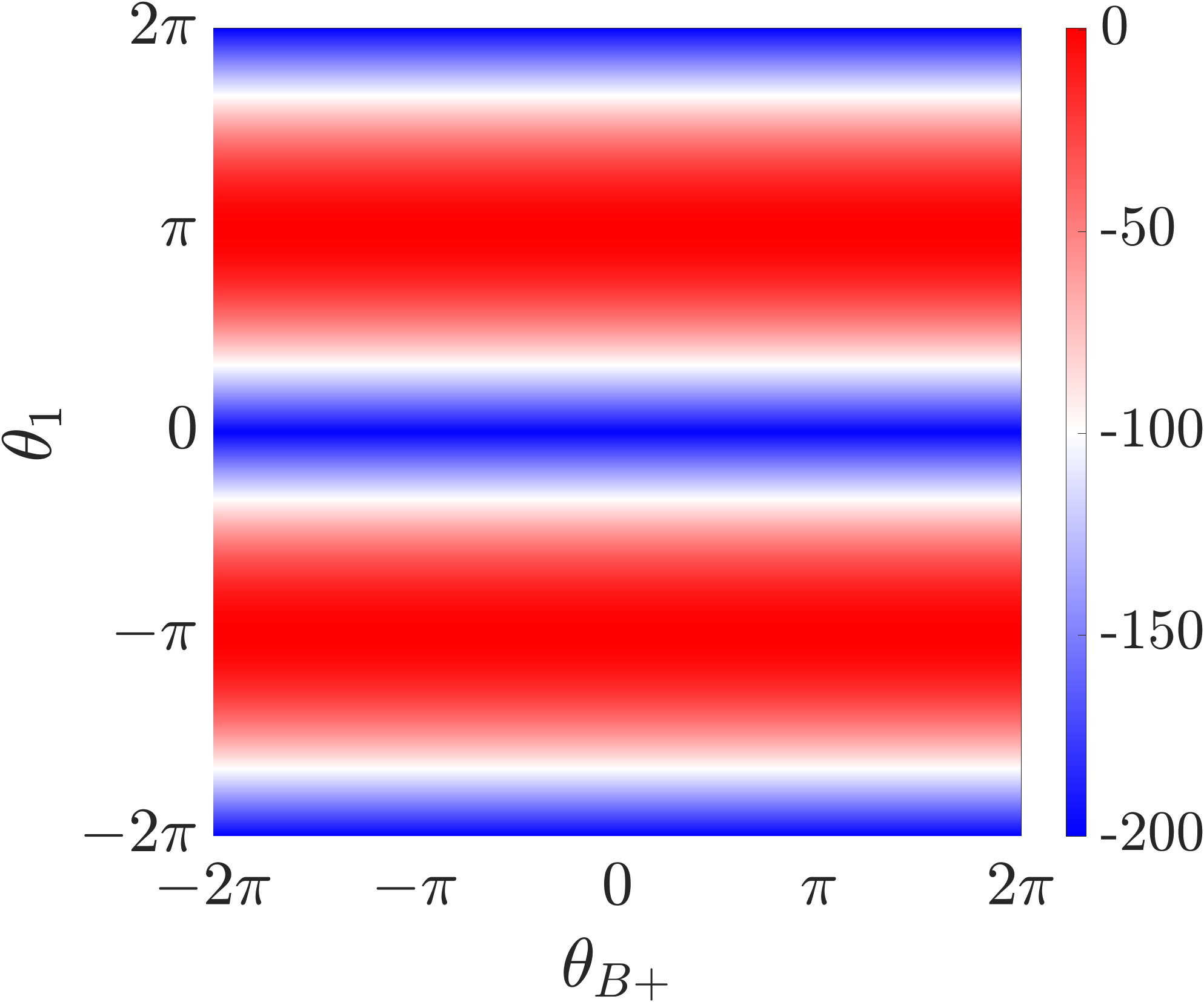}}
	\subfigure[]{\includegraphics[height=0.20\textwidth]{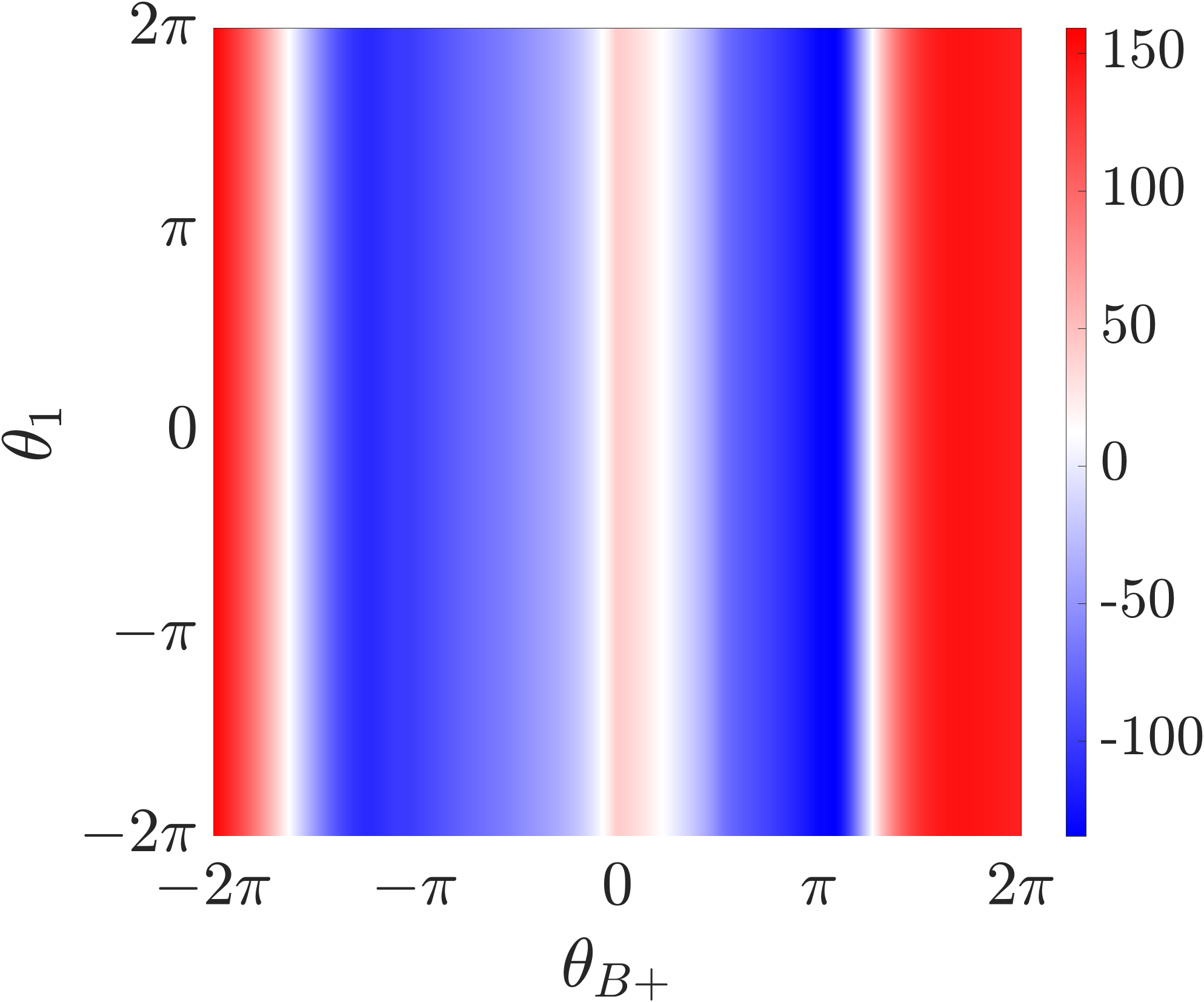}}
	\subfigure[]{\includegraphics[height=0.20\textwidth]{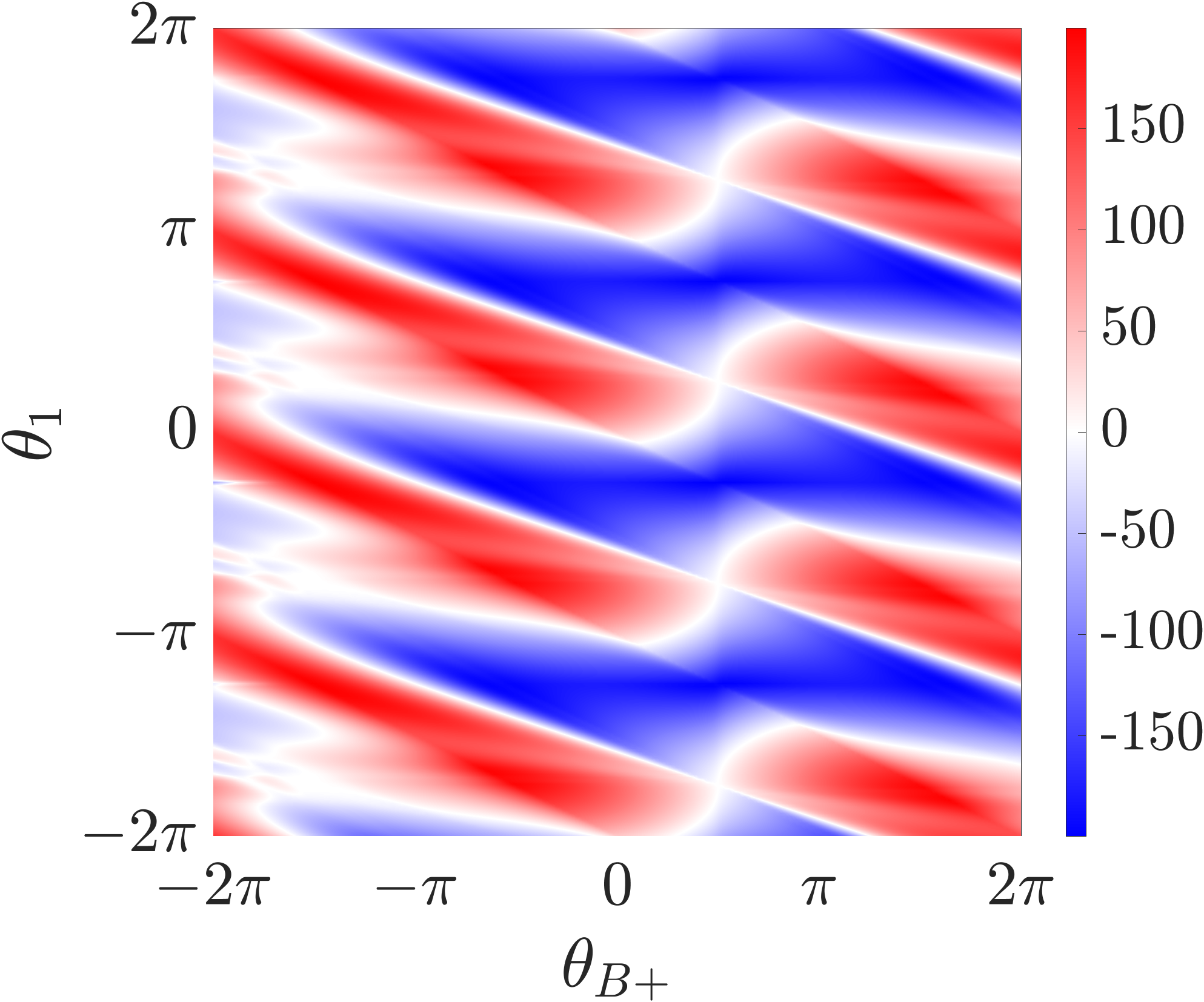}}
	\subfigure[]{\includegraphics[height=0.20\textwidth]{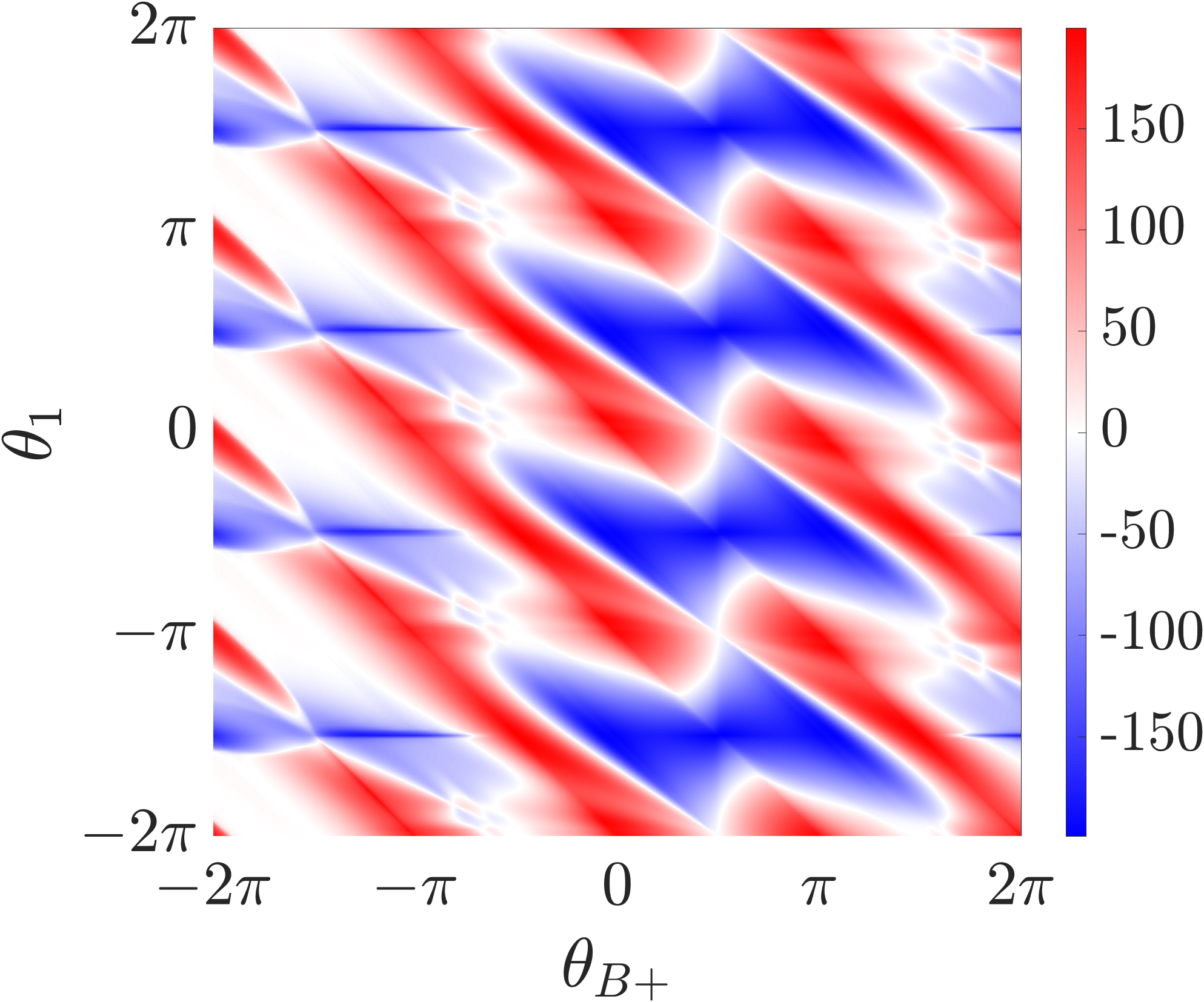}}
	\caption{The expectation value of the position of the walker as a function of the coin parameters $\theta_A$ and $\theta_{B+}$ for the individual strategies (a) \Coin{A}, (b) \Coin{B} and the combined sequences (c) $\mathcal{C}_{AB}^{(2,1)} = \mathcal{C}_A^2 \mathcal{C}_B$, and (d) $\mathcal{C}_{AB}^{(2,2)} = \mathcal{C}_A^2 \mathcal{C}_B^2$ for $t = 200$ time steps. The coin parameter, $\theta_{B-}$ for coin \Coin{B} is kept fixed at $\pi/2$. The \Red{red} and the \Blue{blue} shaded regions again correspond to the \Red{winning} and \Blue{losing} situation, respectively. The system size is taken to be $N = 501$. All the other parameters are the same as in Fig.~\ref{fig:thetapm}.}
	\label{fig:theta1theta2p}
\end{figure*}

Moreover, the choice of the two parameters $(\theta_{B-}, \theta_{B+})$ of $\mathcal{C}_B$ is not restricted to the one which is used till so far and reported in Fig.~\ref{fig:site-dependent}. We investigate the full phase space comprising of three variables at hand, namely, $\theta_A$, $\theta_{B-}$, and $\theta_{B+}$. To start with, we take a fixed value of $\theta_A$ and plot the probability bias (expectation value of the position operator) as a function of other parameters $\theta_{B-}$ and $\theta_{B+}$, for individual and for different combined strategies, in Fig.~\ref{fig:thetapm}. In Fig.~\ref{fig:theta1theta2p}, we plot the expectation value of the position operator as a function of $\theta_A$ and $\theta_{B+}$ by keeping the $\theta_{B-}$ fixed. The plot uses a color gradient to highlight the regions of winning (\Red{red}) and losing (\Blue{blue}) strategies. These plots provide a clear and intuitive visualization of how different parameters of coin operators influence the dynamics of the quantum walk. The red-shaded regions highlight the parameters of the coins where the combined effects of the coin operators result in constructive interference of positive drifting dynamics, thereby producing a net positive drift in the expectation value. These parameters demonstrate the manifestation of Parrondo's paradox, where individually losing strategies combine to yield a winning outcome.

Apart from Ref.~\cite{Kitagawa2012}, where such site dependence is encoded in coin operators in photonic settings, recently, a study has been reported to construct quantum circuits that realize discrete-time quantum walks with an arbitrary position-dependent coin operator~\cite{Nzongani2023}.

We explore further a probabilistic scheme in Appendix~\ref{appen:probparr} where the qualitative behavior remains the same, and we have a winning situation using the combined strategies with appropriate weights for the coins.

Note, we have also considered $\mathcal{C}_{AB}^{(1,1)}$ = \Coin{A}\Coin{B} and $\mathcal{C}_{AB}^{(1,2)}$ = \Coin{A}$\mathcal{C}_B^2$; however, these two sequences do not provide any advantage over the individual coin operators which are evident from the results in Fig.~\ref{fig:site-dependent}. The probability current remains biased towards negative $x$-direction with time.

\section{Time-Dependent Coin}
\label{sec:time-dependent}
\begin{figure}
	\centering
	\subfigure[]{\includegraphics[height=0.21\textwidth]{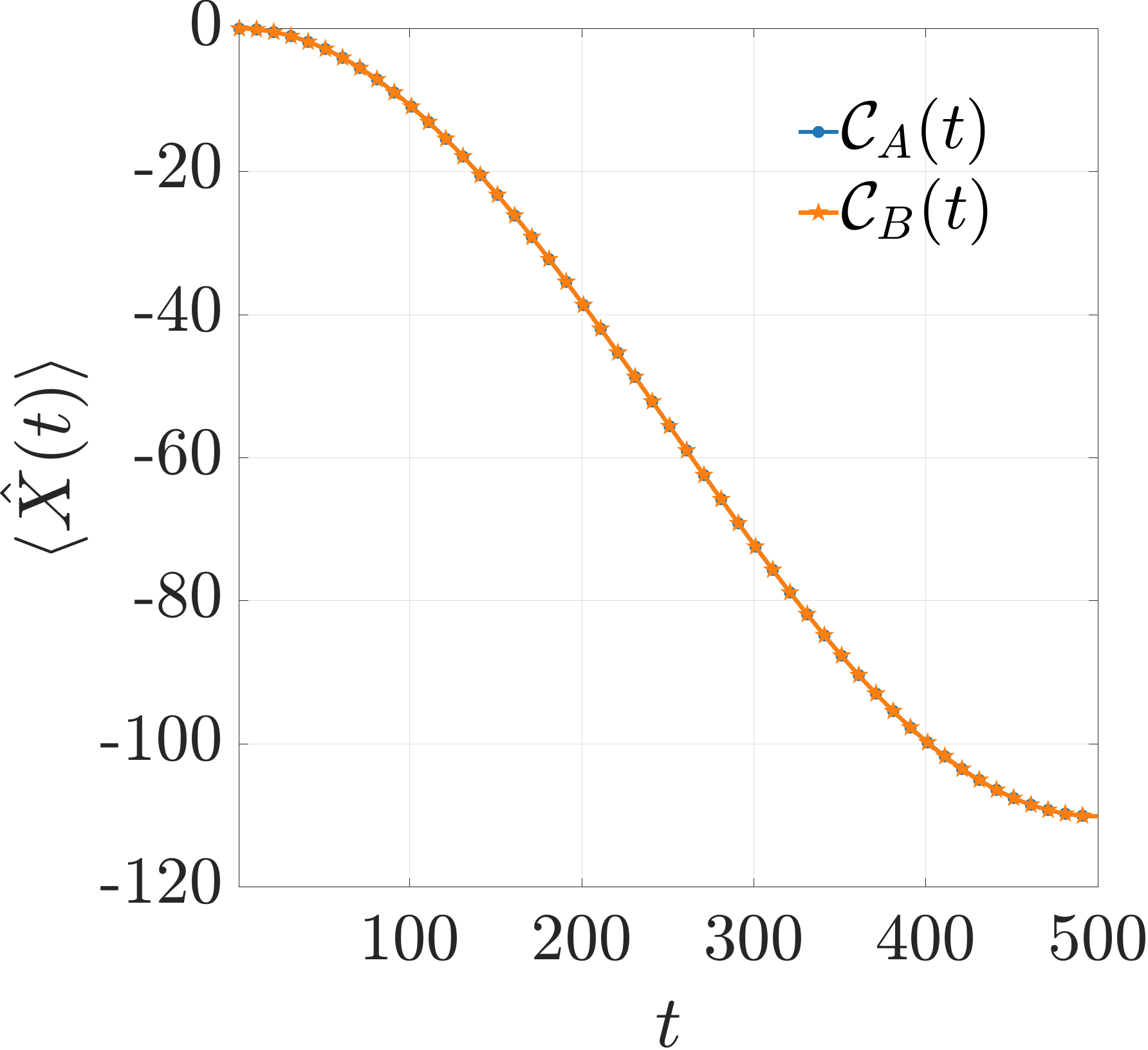}}
	\subfigure[]{\includegraphics[height=0.21\textwidth]{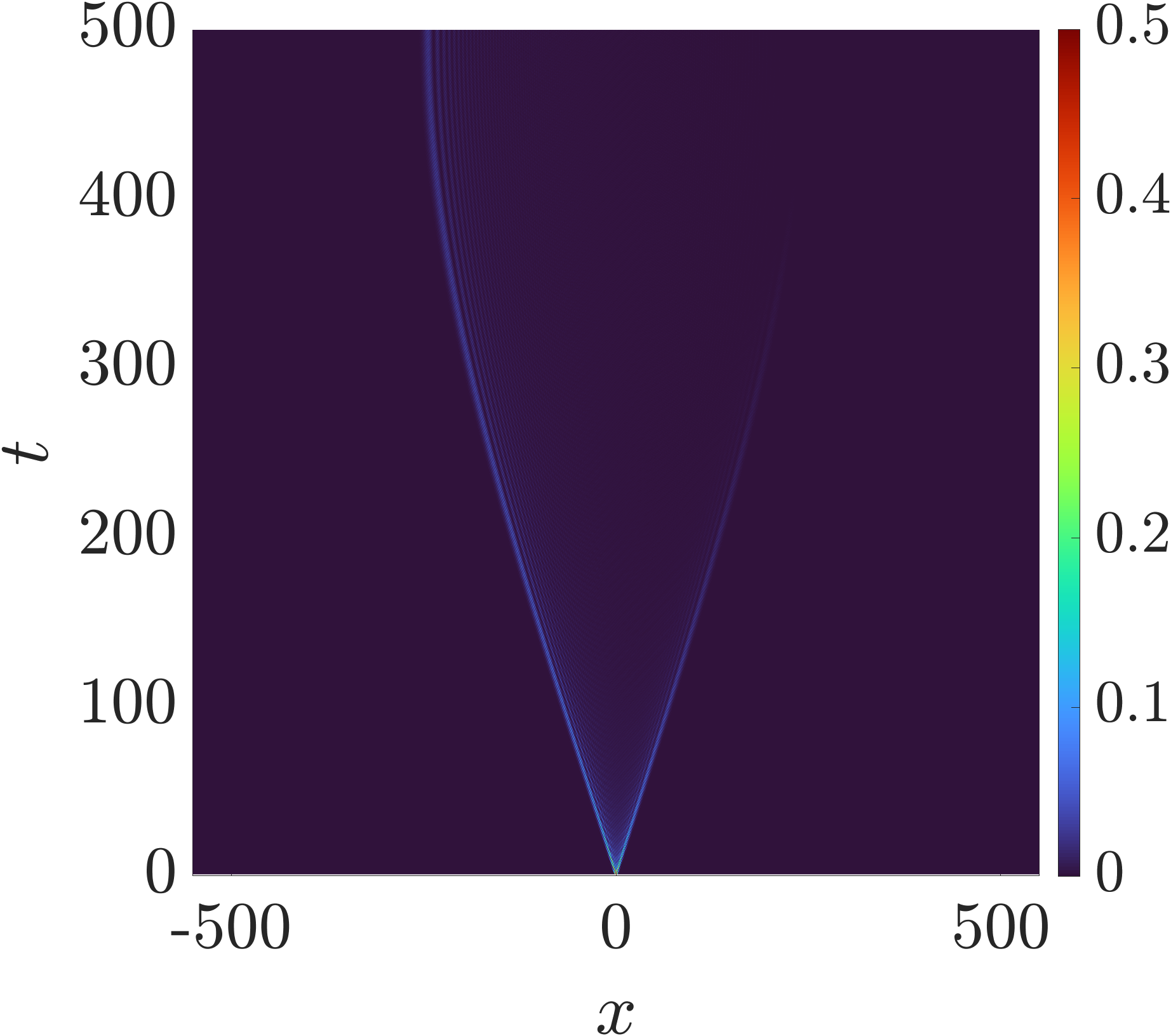}}
	\subfigure[]{\includegraphics[height=0.21\textwidth]{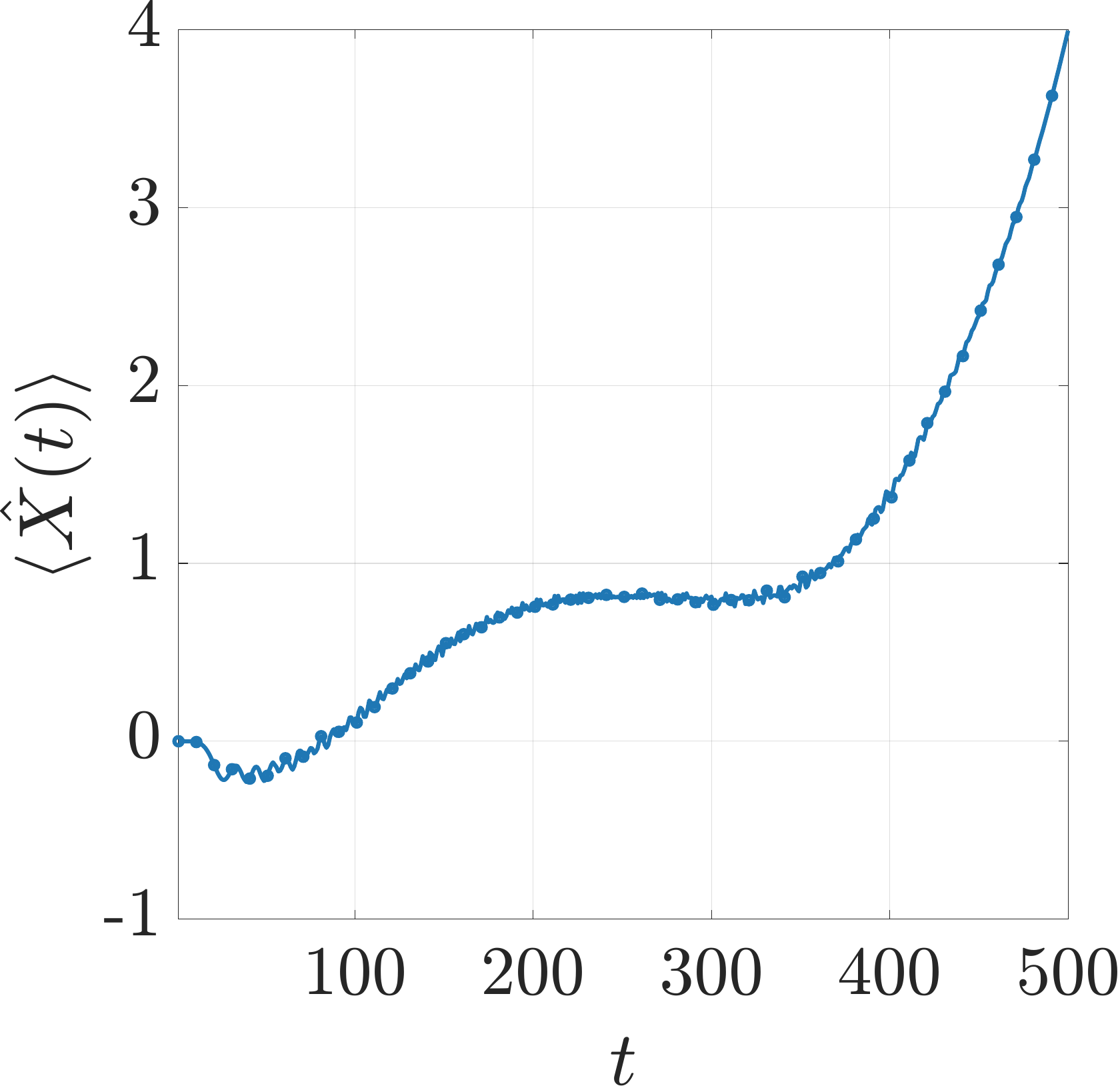}}
	\subfigure[]{\includegraphics[height=0.21\textwidth]{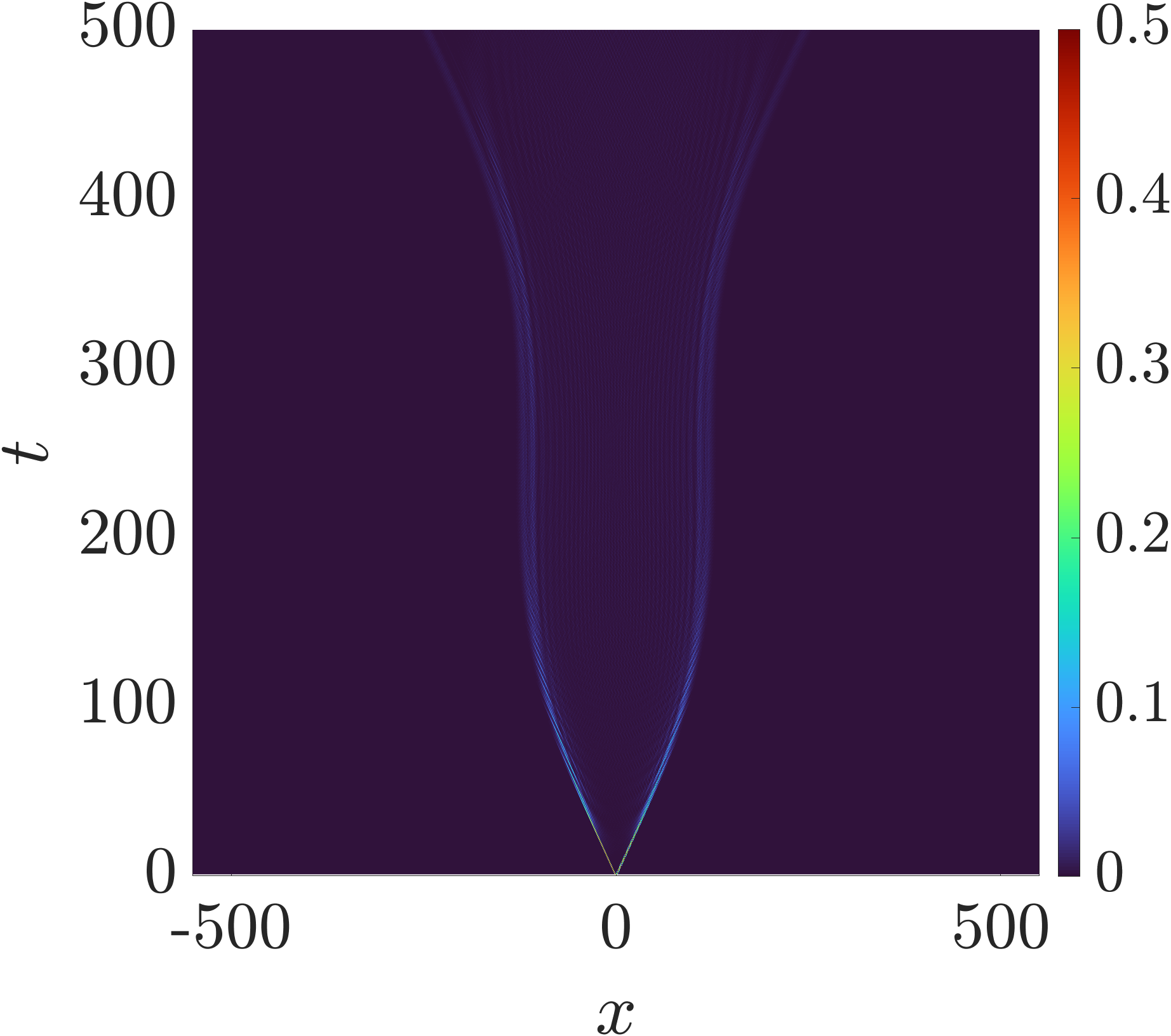}}
    \caption{The probability distribution and the expectation value of the position of the walker as a function of time for different sequences of coins. The two different coins $\mathcal{C}_A(t)$ and $\mathcal{C}_B(t)$ are characterized by $\alpha(t)$ and $\beta(t)$ respectively. In the top row, we plot the dynamics for the individual coins $A$ and $B$. In the bottom row, we consider a combined strategy, \Coin{AB}(t), given by Eq.~\eqref{eq:stepcomb} where $\mathcal{C}_A(t)^2$ is used for an even number of time steps and $\mathcal{C}_B(t)^2$ otherwise. The system size is taken to be $N = 1001$. The composite system is initialized in $\ket{\Psi(0)} = \ket{\psi}_C \otimes \ket{\psi}_L = (\ket{\uparrow} + i \ket{\downarrow})/\sqrt{2} \otimes \ket{0}$.}
	\label{fig:time-dependenttrivial}
\end{figure}

In Ref.~\cite{Sanders2015}, the authors demonstrated a quantum walk with a time-dependent coin bias in a photonic setup. By linearly ramping the time-dependent coin flip operation, they observed periodic revivals of the walker distribution. This work sheds light on practical implementations of time-dependent coins in quantum walk setups. These experimental studies motivate us to further explore Parrondo's paradox-like scenarios with time-dependent coins. 

We start by considering the following parameterization of the coin operator 
\begin{equation}
	\mathcal{C}(q(t), \alpha(t), \beta(t)) = \begin{pmatrix}
		\sqrt{q(t)} & \sqrt{1 - q(t)} e^{i \alpha(t)} \\
		\sqrt{q(t)} e^{i \beta(t)} & -\sqrt{q(t)} e^{i (\alpha(t) + \beta(t))}
	\end{pmatrix}
\end{equation}
where the three coin parameters can, in general, be a function of position $x$ or time $t$ or both, depending upon the setup under consideration. For our purpose, we consider these parameters to be a function of time $t$ only. These parameters are defined as follows: $0 \le q(t) \le 1$, and $0 \le \alpha(t), \beta(t) \le 2 \pi$. For example, it is straightforward to recover some of the common coins, like the Hadamard and Fourier coins, by setting
\begin{equation}
	\mathcal{C}_H(q(t) = 1/2, \alpha(t) = 0 = \beta(t)) = \dfrac{1}{\sqrt{2}} \begin{pmatrix}
		1 & 1 \\
		1 & -1
	\end{pmatrix}
\end{equation}
and 
\begin{equation}
	\mathcal{C}_F(q(t) = 1/2, \alpha(t) = \pi/2 = \beta(t)) = \dfrac{1}{\sqrt{2}} \begin{pmatrix}
		1 & i \\
		i & 1
	\end{pmatrix}
\end{equation}
respectively. For our purpose, we start a similar game where the two players, $A$ and $B$, are provided with two different coins $\mathcal{C}_A(t)$ and $\mathcal{C}_B(t)$, which are characterized as follows:
\begin{equation}
	\text{$\mathcal{C}_A(t)$} = \mathcal{C}(1/2, \alpha(t), 0), \;\;\;  \text{$\mathcal{C}_B(t)$} = \mathcal{C}(1/2, 0, \beta(t)).
\end{equation}
with
\begin{equation}
    \alpha(t) = \dfrac{2 \pi}{T} t = \beta(t)
\end{equation}
where $T$ is the total number of time steps. In Fig.~\ref{fig:time-dependenttrivial}, we plot the expectation value of the position operator and the probability distribution for the walker as a function of time steps for the individual coins given to the two players. As the quantum walk evolves, the behavior is similar to the case of a site-dependent coin, and we do not see any revival in the dynamics. Hence, it is not possible to win the game with individual coins. In this case, the composite system is initialized in the following state
\begin{equation}
	\ket{\Psi(0)} = \dfrac{1}{\sqrt{2}} (\ket{\uparrow} + i \ket{\downarrow}) \otimes \ket{0}.
\end{equation}

\subsection*{Winning Strategy}
Now, we introduce one of the strategies or a sequence of coin operators, which is a combination of the individual coins $\mathcal{C}_A(t)$ and $\mathcal{C}_B(t)$ such that it results in a probability distribution with a bias towards the right to the origin. It is given as 
\begin{equation}
	\mathcal{C}_{AB}(t) = \begin{cases}
		\mathcal{C}_{A}(t)^2	, & \text{if} \;\; t \in \text{even} \\
		\mathcal{C}_{B}(t)^2, & \text{if} \;\; t \in \text{odd} \\
	\end{cases}
 \label{eq:stepcomb}
\end{equation}
i.e., we use the coins twice and alternatively in order to win the game. We now choose the above coin operator to evolve the quantum walk and plot the behavior of the expectation value of the position and the probability distribution as a function of time steps $t$ in Fig.~\ref{fig:time-dependenttrivial}. We observe the shift of probability current towards the right as the quantum walk evolves. 
Such strategies~\cite{Duarte2020}, which involve alternate actions of different coins, have been reported recently to recreate Parrondo's paradox, although with different choices of coins.

\section{Conclusion}
\label{sec:conclusion}
In conclusion, our investigation into Parrondo's paradox in discrete-time quantum walks reveals that the paradox can be induced using site- and time-dependent coin operators without the need for multi-state coins or decoherence. This finding challenges the conventional understanding and demonstrates that a simpler, more elegant approach can achieve the same paradoxical outcomes previously reported in more complex setups. Unlike earlier methods that introduced additional layers of complexity, our use of general SU(2) rotations for the coin operators provides greater flexibility and control over the quantum walk dynamics, making it more feasible for practical implementation and experimental verification. Our results not only simplify the experimental realization of Parrondo's paradox but also open up new avenues for exploring quantum strategies and their applications in quantum transport, computation, and information processing. This study paves the way for more accessible and versatile quantum algorithms, contributing significantly to the advancement of quantum technologies. It would be intriguing to see if Parrrondo's paradox can be generalized for more than two games. In the present study, we considered a handful of combined strategies, namely, certain combinations of $m$ and $n$ to realize Parrondo's paradox, and it would be interesting to find out which other combinations of $(m, n)$ leads to a winning situation. This could be a natural extension of the present work for the future.

\section{Acknowledgments}
 This research has been supported by the MOST Young Scholar Fellowship (Grants No. 112-2636-M-007-008- and No. 113-2636-M-007-002-), National Center for Theoretical Sciences (Grants No. 113-2124-M-002-003-) from the Ministry of Science and Technology (MOST), Taiwan, and the Yushan Young Scholar Program (as Administrative Support Grant Fellow) from the Ministry of Education, Taiwan.

\clearpage

\appendix

\section{Probabilistic Parrondo sequence}
\label{appen:probparr}
\begin{figure*}
    \centering
    \subfigure{\includegraphics[height=0.22\textwidth]{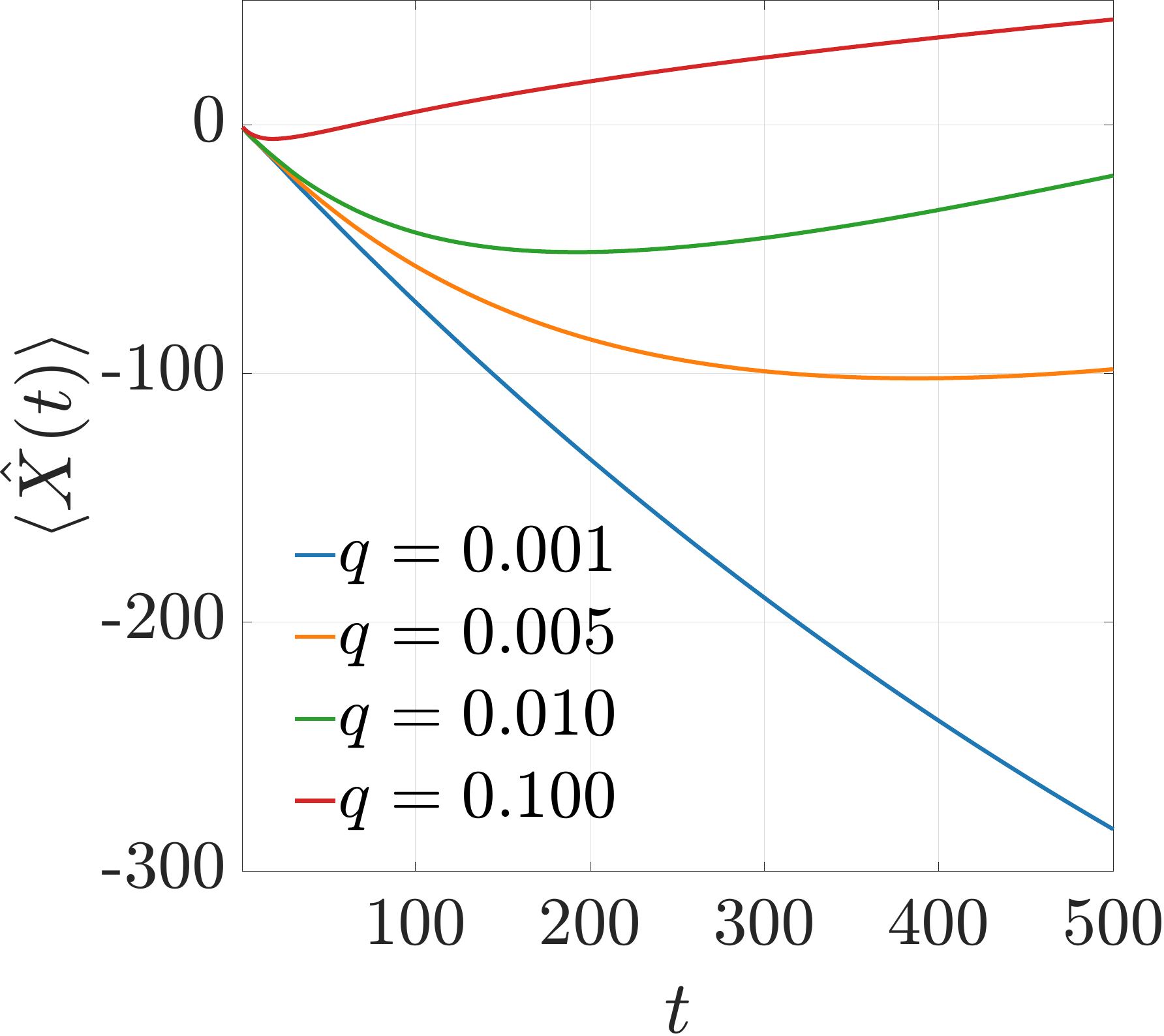}}
    \subfigure{\includegraphics[height=0.22\textwidth]{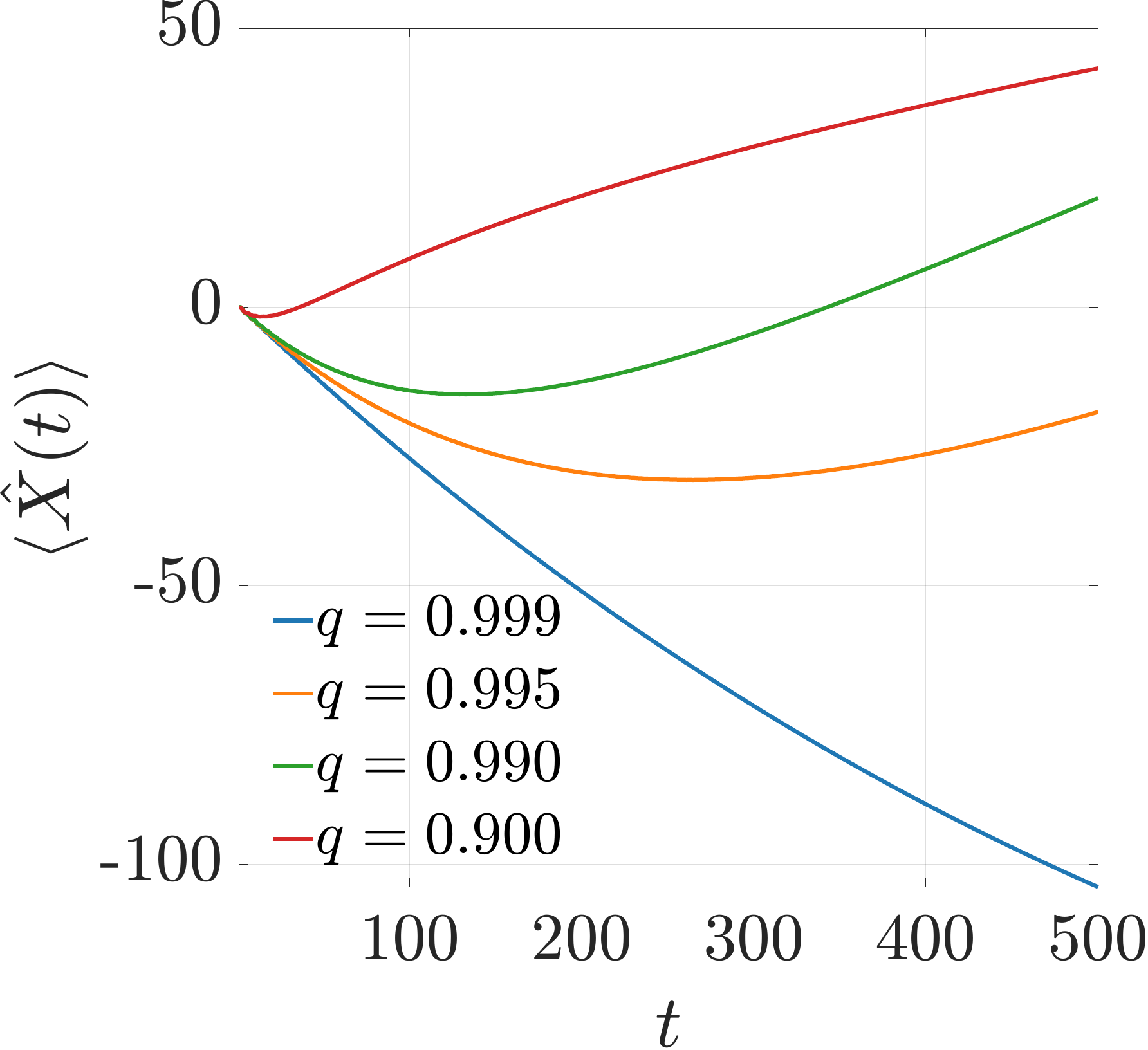}}
    \subfigure{\includegraphics[height=0.22\textwidth]{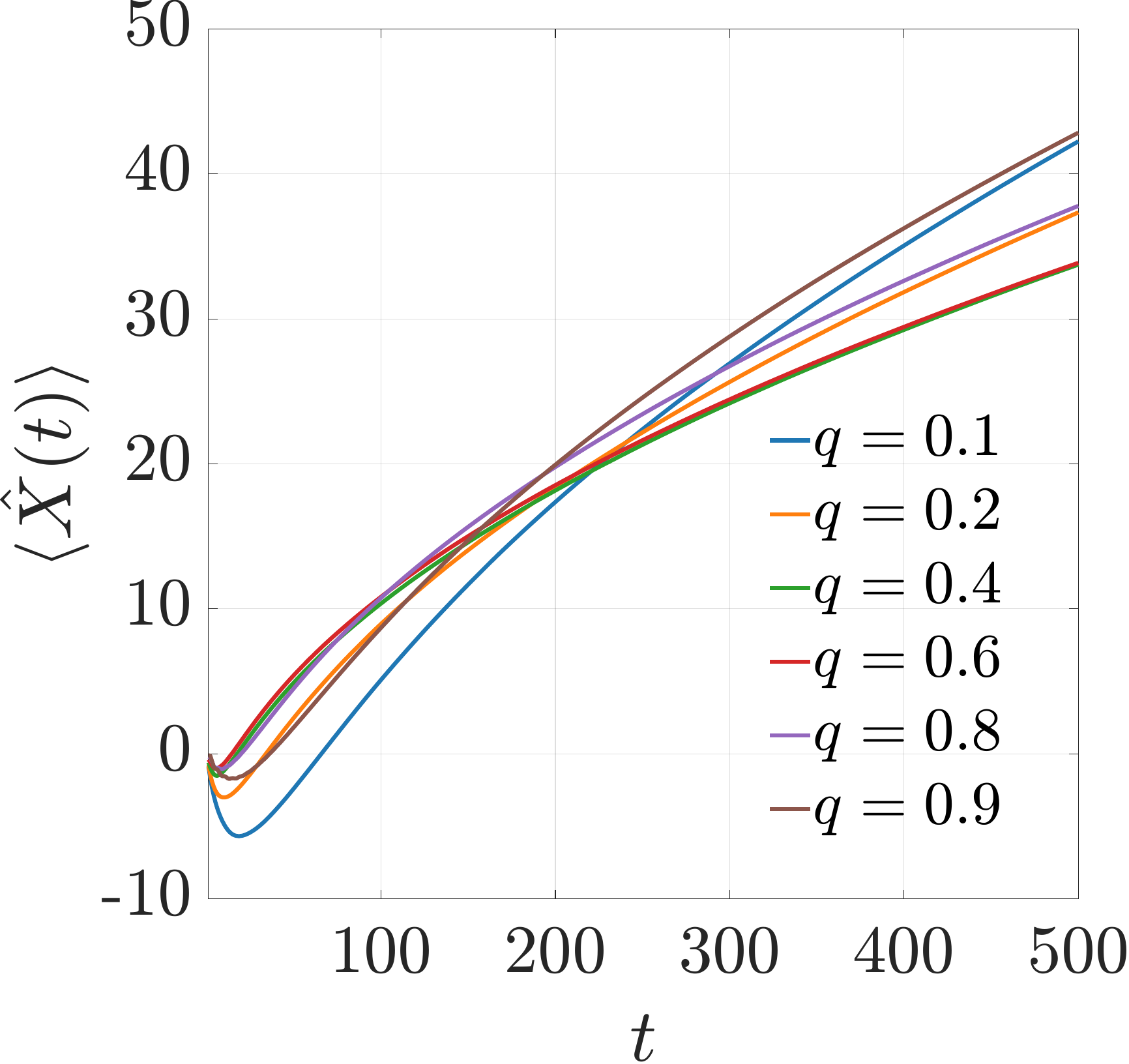}}
    \caption{The expectation value of the position of the walker as a function of time for the combined probabilistic strategy discussed in Appendix~\ref{appen:probparr} and is given in Eq,~\ref{eq:probcomb}. The plots correspond to different values of $q$ and an average is taken over $50000$ iterations. }
    \label{fig:probilistic}
\end{figure*}

While a deterministic sequence of alternating quantum coins can demonstrate Parrondo's paradox, another intriguing approach is to combine these strategies probabilistically. In a probabilistic sequence, the quantum walker switches between different coin operators according to specific probabilities rather than a fixed pattern. This method introduces an additional layer of complexity and randomness, which can still yield a winning outcome despite each strategy being a losing strategy on its own. In a probabilistic sequence, we define a probability $q$ for choosing the coin 
\Coin{A} and a probability $1 - q$ for choosing the coin \Coin{B} at each step, i.e. 
\begin{equation}
    \mathcal{C}_{AB} = q \mathcal{C}_A + (1 - q) \mathcal{C}_B.
    \label{eq:probcomb}
\end{equation}
We plotted the expectation value of the walker's position in Fig.~\ref{fig:probilistic} for different probabilistic strategies and found that Parrondo's effect still occurs for certain coin weights. By adjusting the probabilities, we observed that some combinations led to a net positive movement of the walker, showing a winning situation. This means that even with a mix of strategies, we can still achieve a winning outcome by choosing the right probabilities for the coin flips.

\section{Dependence on the initial state of the coin}
\label{appen:initialstate}
\begin{figure*}
    \centering
    \subfigure[]{\includegraphics[height=0.22\textwidth]{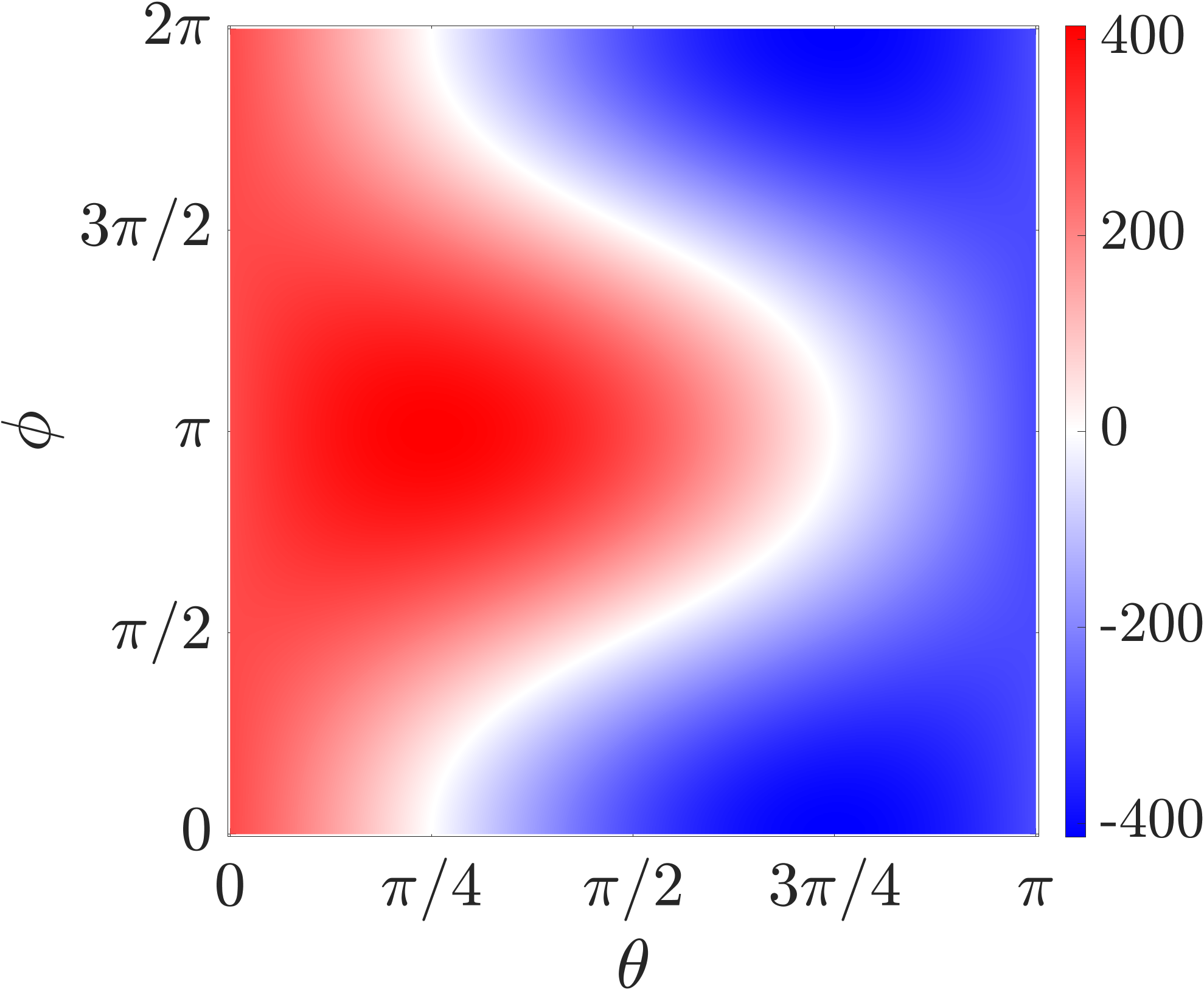}}
    \subfigure[]{\includegraphics[height=0.22\textwidth]{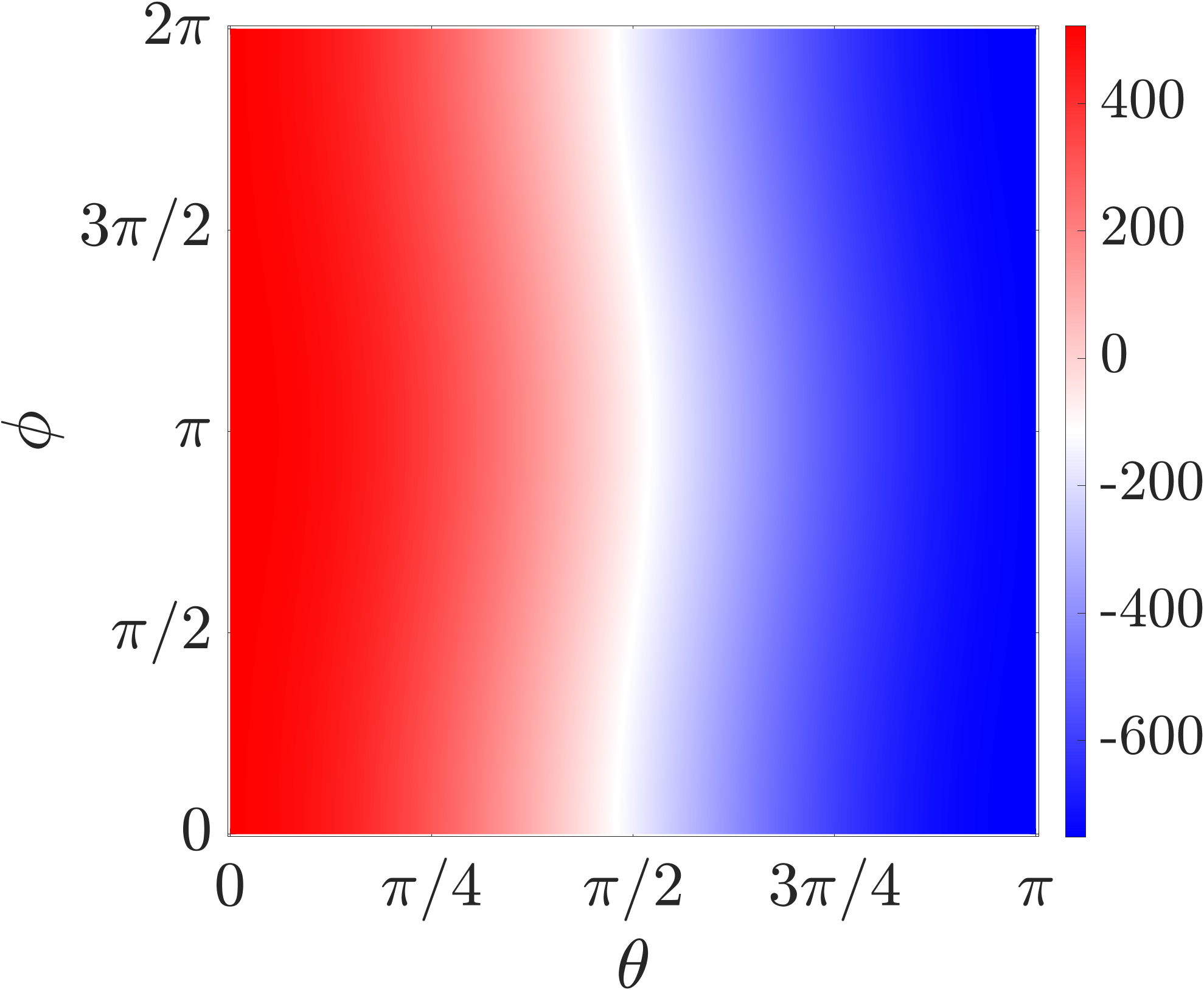}}
    \subfigure[]{\includegraphics[height=0.22\textwidth]{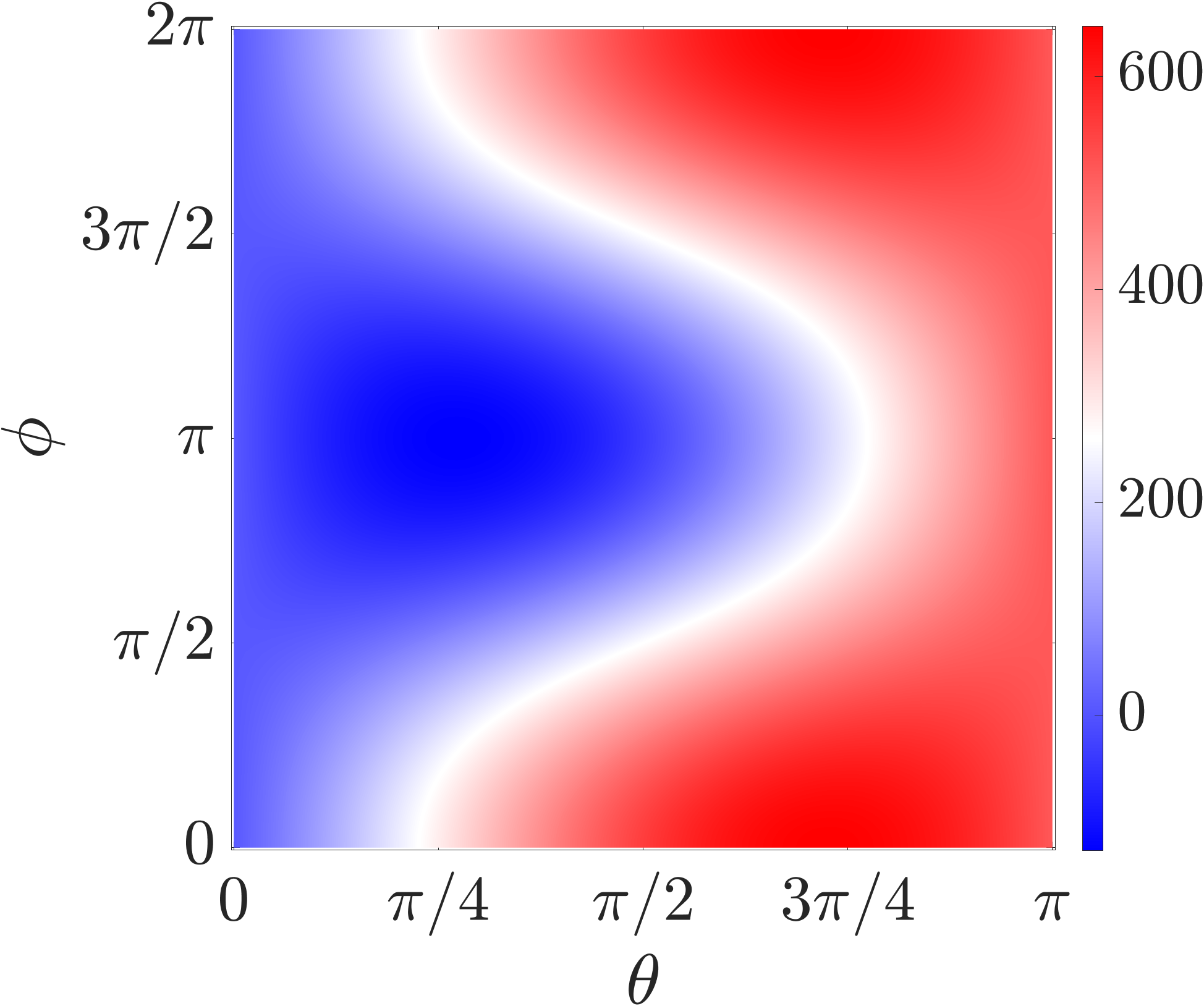}}
    \caption{The expectation value of the position of the walker at a time $t = 1000$ as a function of the parameters of the initial state of the coin $\theta$ and $\phi$ given in Eq.~\ref{eq:initialcoin}. The two individual coins $\mathcal{C}_A$ and $\mathcal{C}_B$ are characterized by $\theta_A = \pi/2$ and $(\theta_{B-}, \theta_{B+}) = (-\pi/8, \pi/4)$ respectively. The coin sequences are chosen to be (a) $\mathcal{C}_A$, (b) $\mathcal{C}_B$ and (c) $\mathcal{C}_{AB}^{(2,2)} = \mathcal{C}_A^2 \mathcal{C}_B^2$. The \Red{red} and the \Blue{blue} shaded regions correspond to the \Red{winning} and \Blue{losing} situation, respectively. The system size is $2101$ and the walker is localized at $\ket{x = 0}$ at $t = 0$.}
    \label{fig:intialstatesite}
\end{figure*}

\begin{figure*}
    \centering
    \subfigure[]{\includegraphics[height=0.22\textwidth]{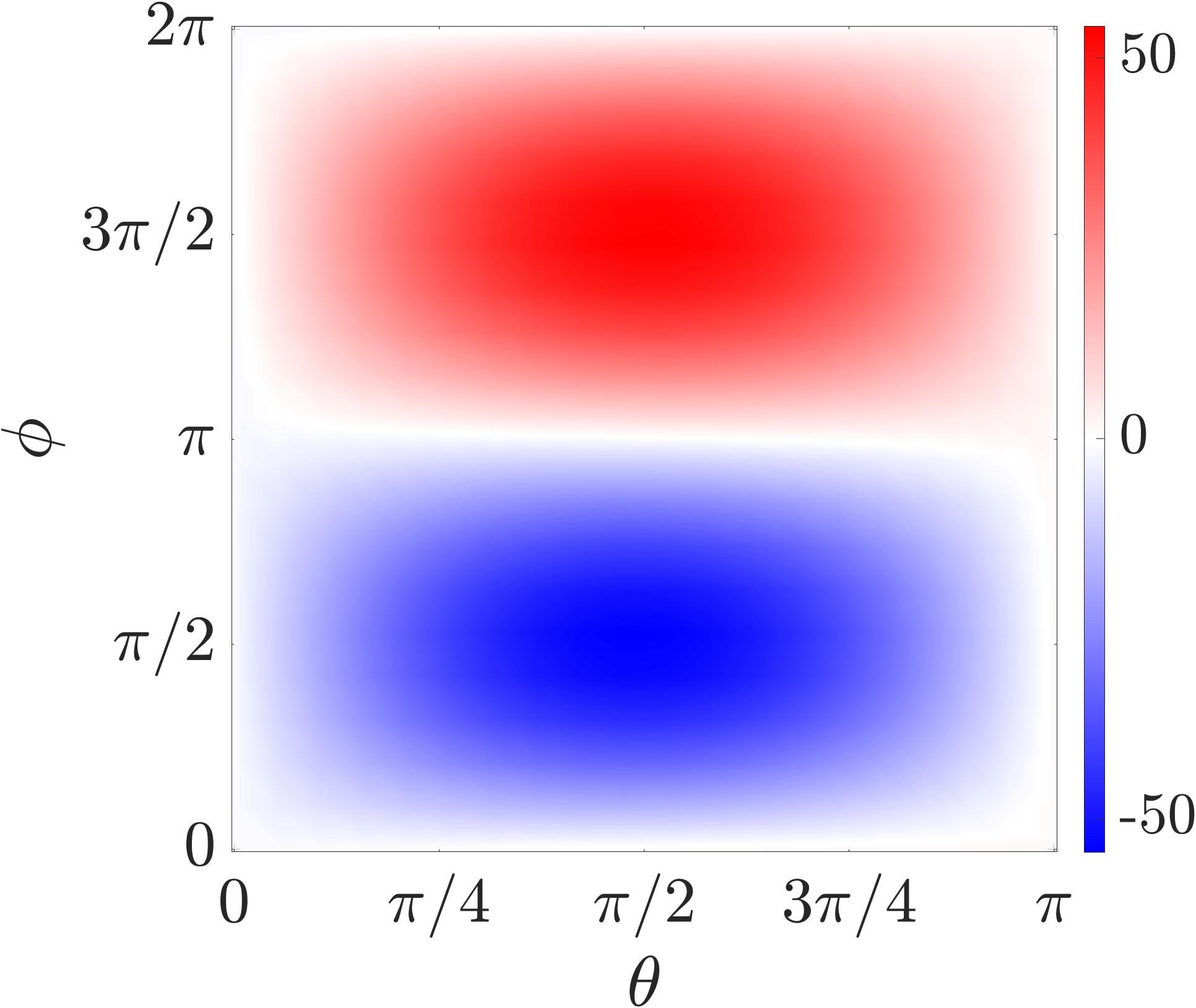}}
    \subfigure[]{\includegraphics[height=0.22\textwidth]{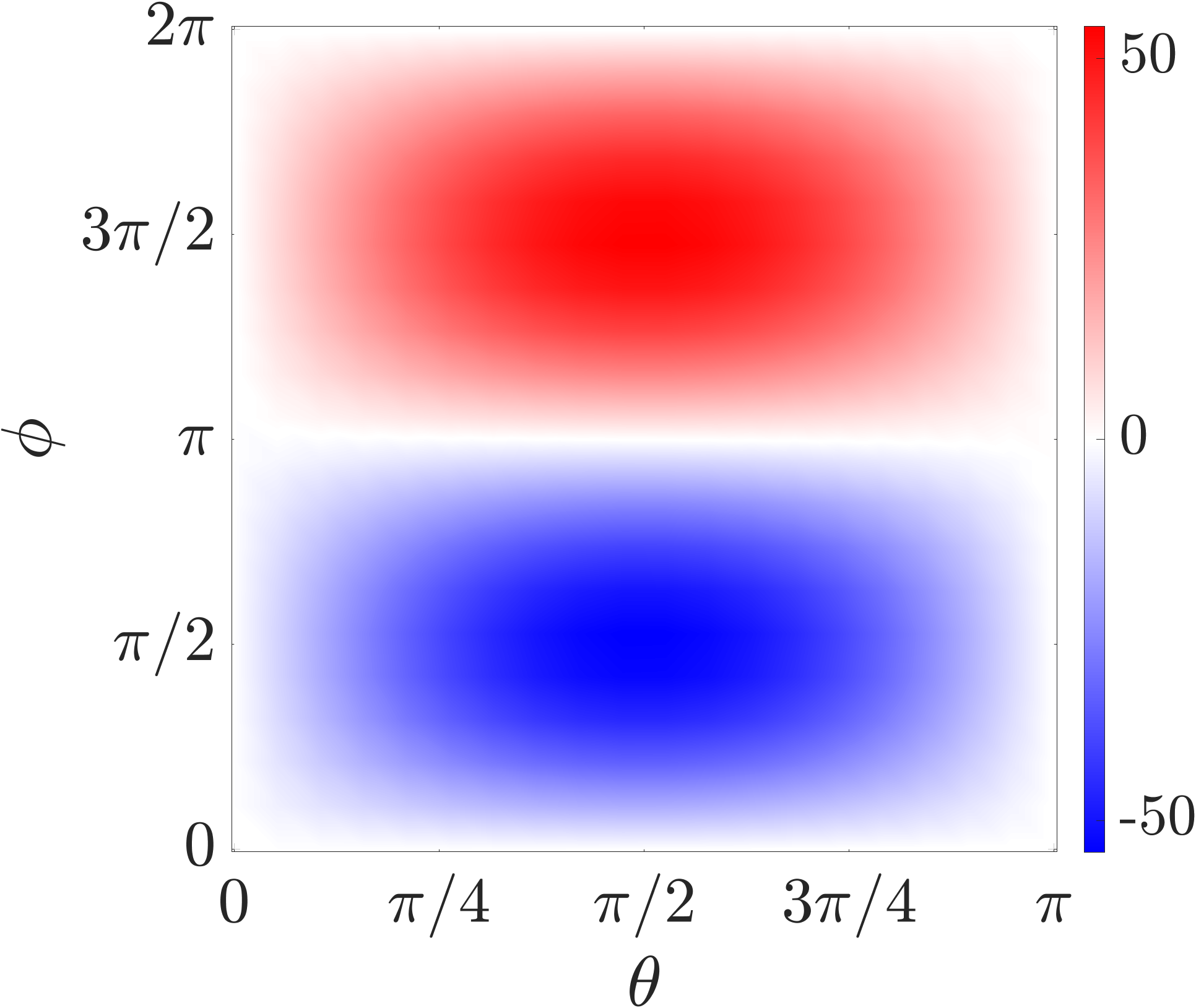}}
    \subfigure[]{\includegraphics[height=0.22\textwidth]{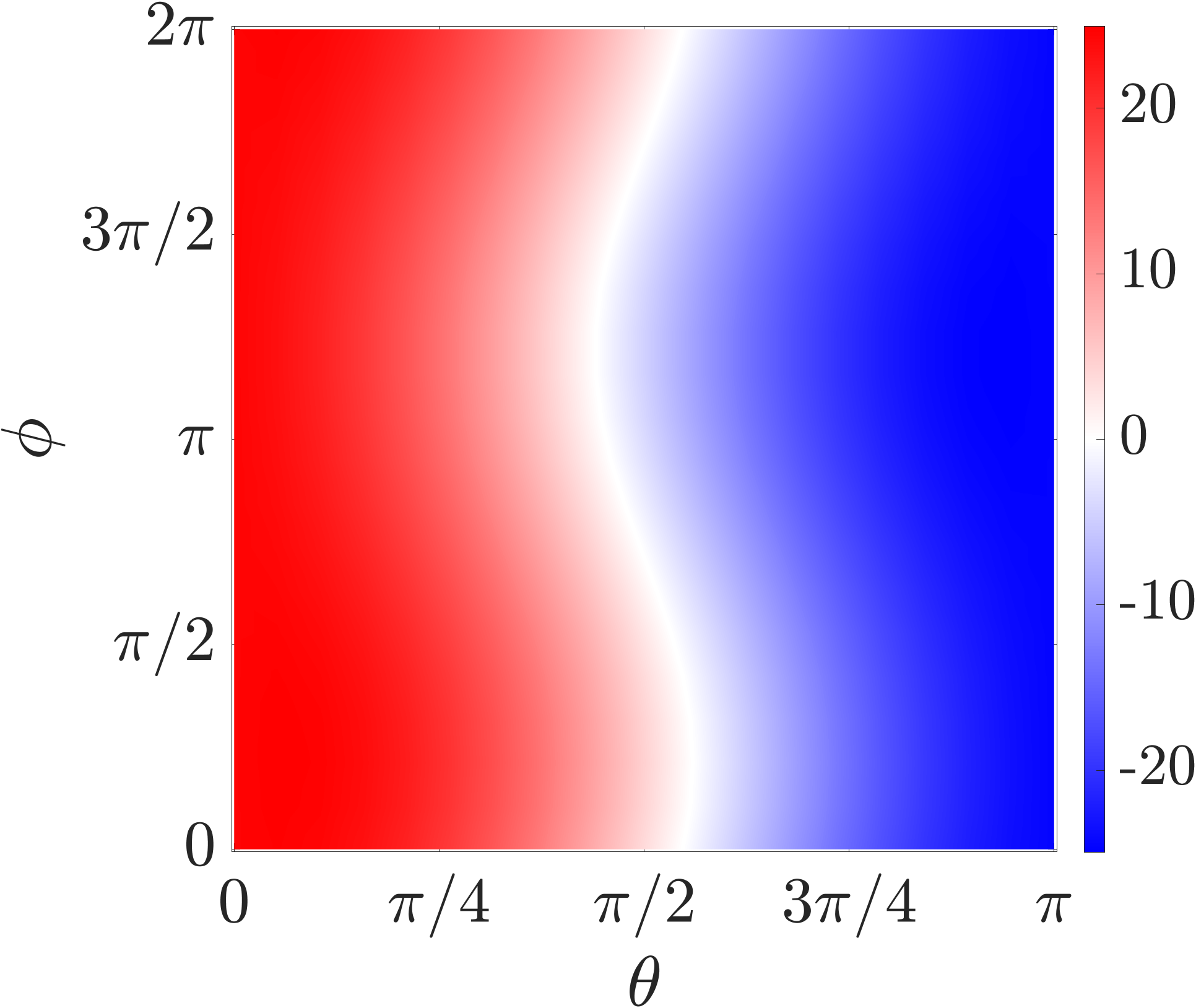}}
    \caption{The expectation value of the position of the walker at a time $t = 250$ as a function of the parameters of the initial state of the coin $\theta$ and $\phi$ given in Eq.~\ref{eq:initialcoin}. The two individual coins $\mathcal{C}_A$ and $\mathcal{C}_B$ are characterized by $C(1/2, \alpha(t), 0)$ and $C(1/2, 0, \beta(t))$ respectively. The coin sequences are chosen to be (a) $\mathcal{C}_A$, (b) $\mathcal{C}_B$, and (c) $\mathcal{C}_A^2$ is used for an even number of time steps and $\mathcal{C}_B^2$ for odd one. The \Red{red} and the \Blue{blue} shaded regions correspond to the \Red{winning} and \Blue{losing} situation, respectively. The system size is $401$ and the walker is localized at $\ket{x = 0}$ at $t = 0$.}
    \label{fig:initialstatestep}
\end{figure*}

We explore the role of the initial state of the coin on the manifestation of Parrondo's paradox in discrete-time quantum walks and report significant insights. We found that the initial coin state plays a crucial role in determining the dynamics and eventual outcomes of the quantum walk. Specifically, a most general state of the coin, which is a superposition of the basis state $\ket{\uparrow}$ and $\ket{\downarrow}$, characterized by the parameters $\theta$ and $\phi$, given by
\begin{equation}
    \label{eq:initialcoin}
    \ket{\psi}_C = \cos(\dfrac{\theta}{2})\ket{\uparrow} + e^{i \phi} \sin(\dfrac{\theta}{2})\ket{\downarrow},
\end{equation}
directly influence the interference patterns and probability distributions that emerge during the walk. By systematically varying the parameters $\theta$ and $\phi$, and hence the initial state of the coin, we plotted the expectation value of the walker's position using site-dependent and step-dependent coins for different initial states in Fig.~\ref{fig:intialstatesite} and Fig.~\ref{fig:initialstatestep}, respectively. We observed that certain configurations either enhance or suppress the paradoxical effect. These results show that the choice of initial state and coin configuration can significantly impact the outcome, either strengthening or weakening the paradoxical behavior.

This sensitivity to the initial coin state underscores the intricate interplay between quantum coherence, superposition, and the probabilistic nature of quantum walks, highlighting the need for precise control and preparation of initial states in experimental realizations. Our findings suggest that by carefully selecting the initial coin state, one can optimize the conditions for observing Parrondo's paradox, thereby enhancing the practical applicability of quantum walks.

\section{Dynamics for a first few steps}
\label{appen:dynamics}
\begin{figure*}
    \centering
    \subfigure[]{\includegraphics[width=0.3\textwidth]{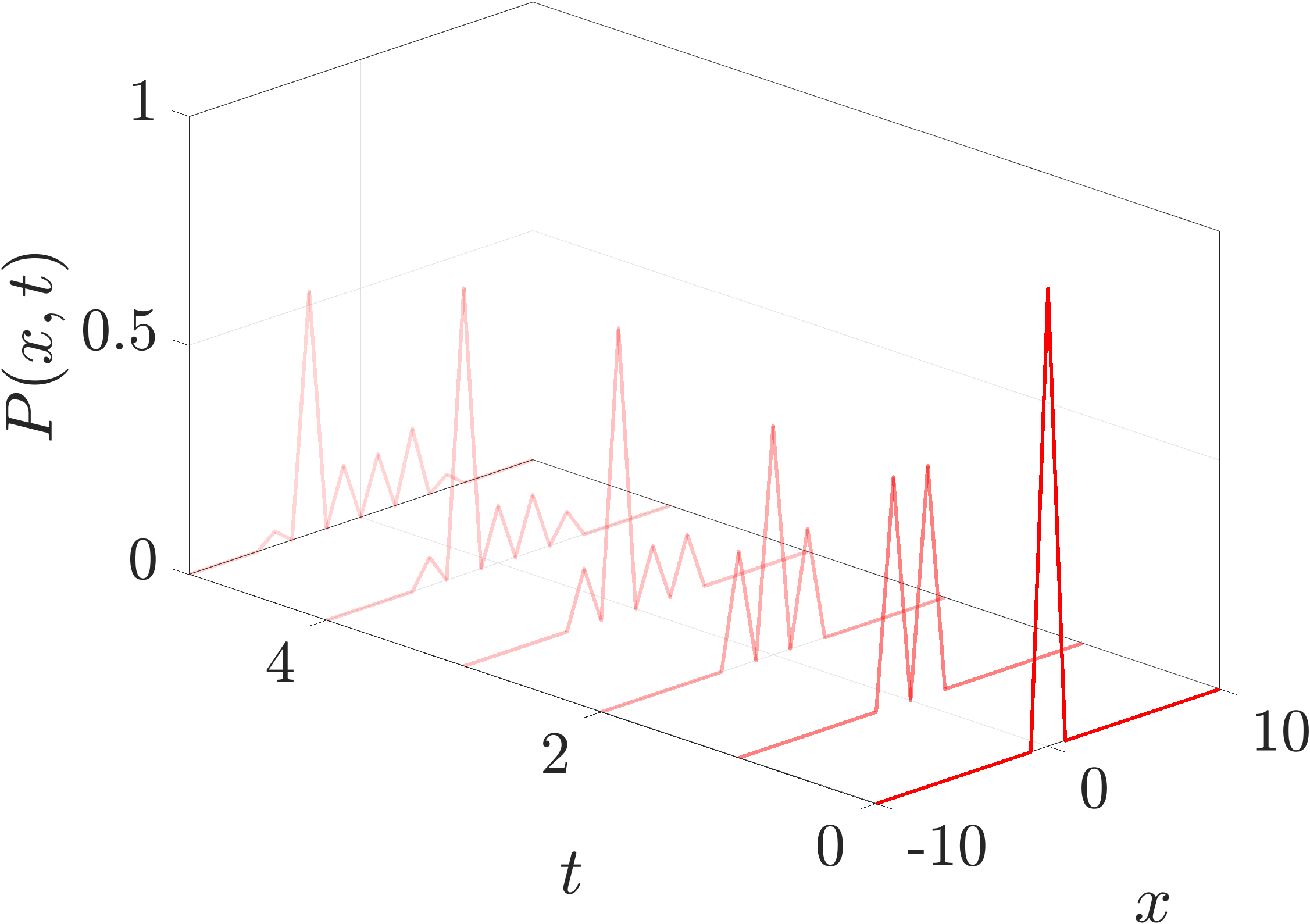}}
    \subfigure[]{\includegraphics[width=0.3\textwidth]{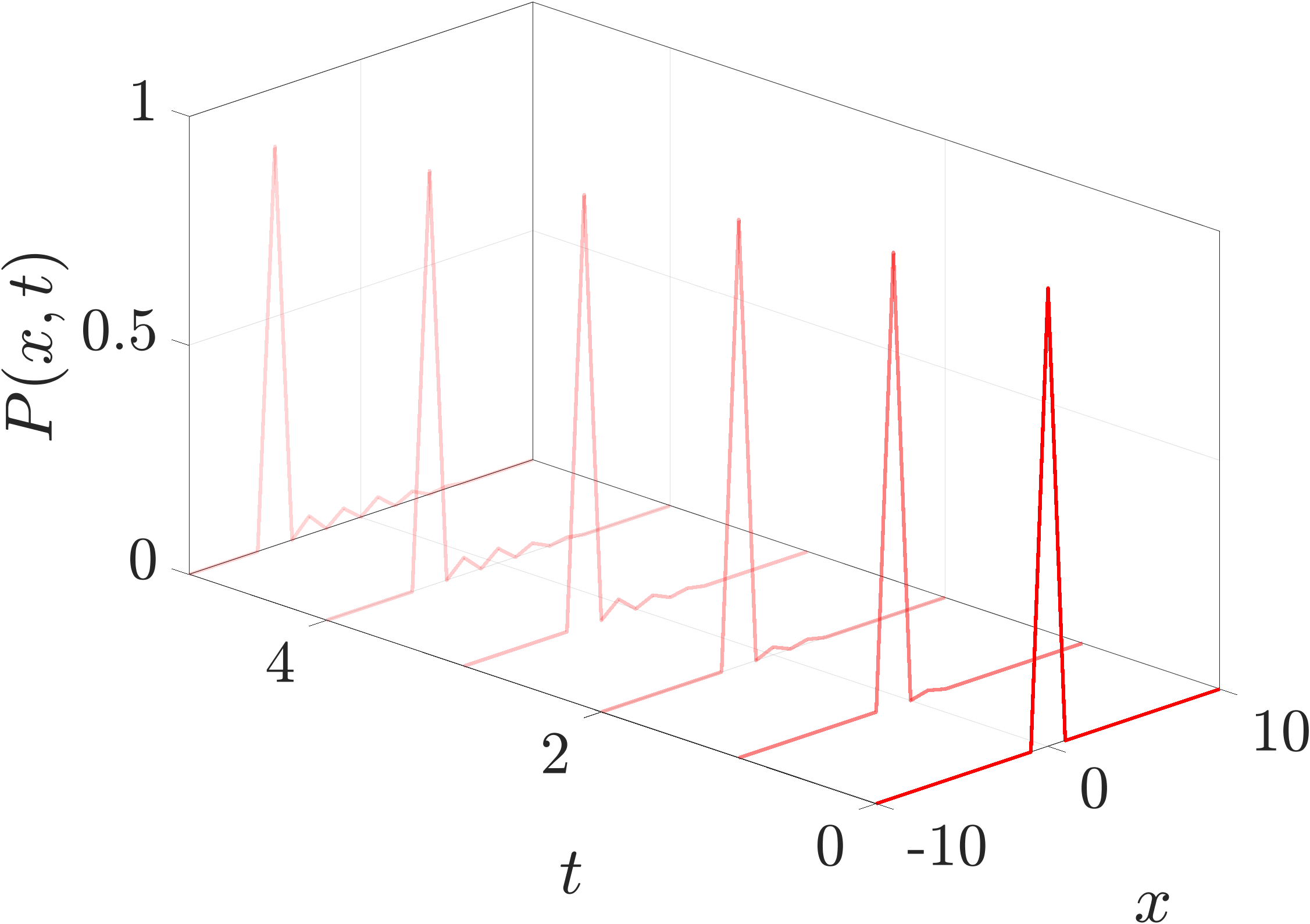}}
    \subfigure[]{\includegraphics[width=0.3\textwidth]{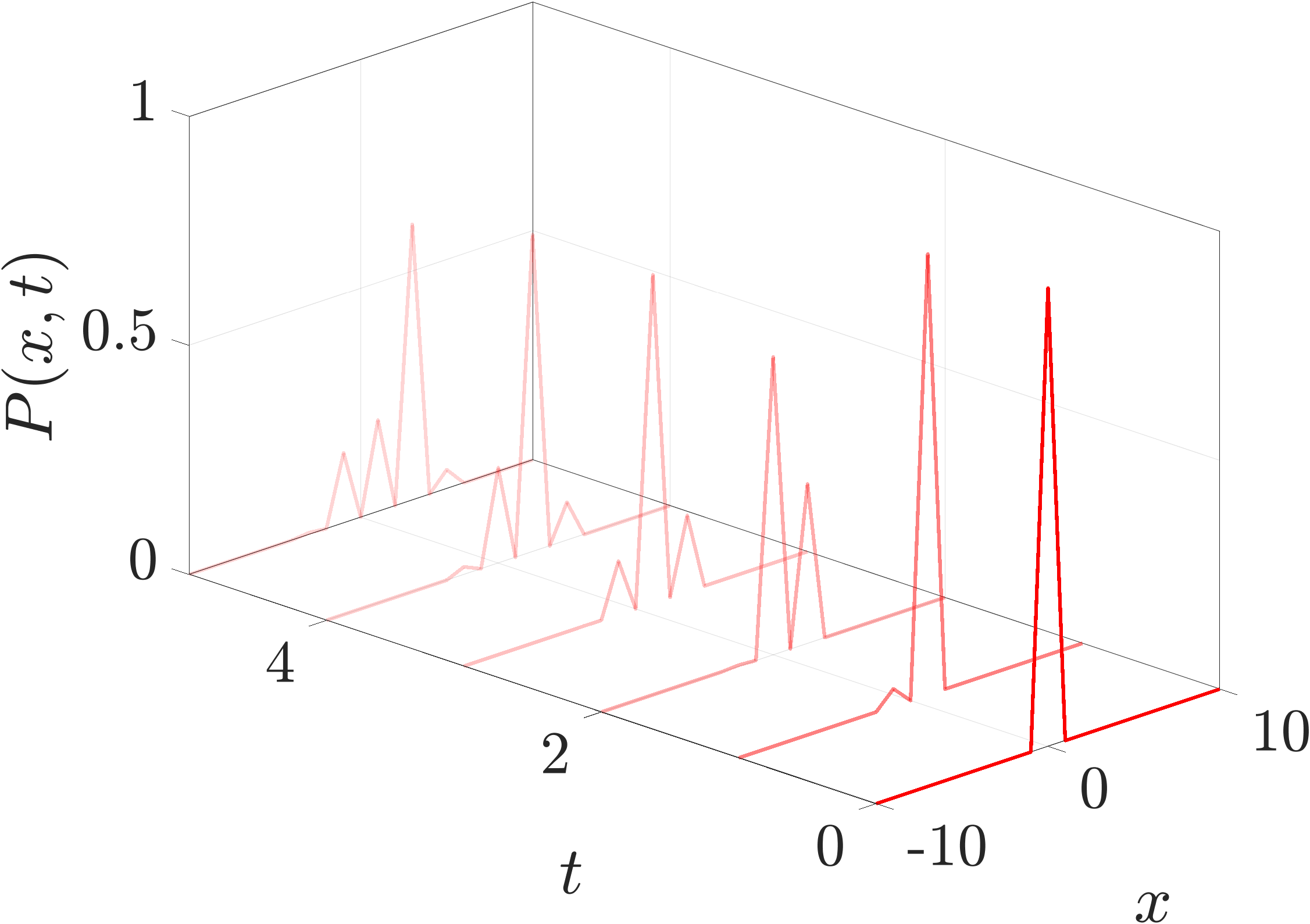}}
    \caption{The probability distribution of the walker for the first few steps for the coins (a) $\mathcal{C}_A$, (b) $\mathcal{C}_B$, and (c) $\mathcal{C}_{AB}^{(2,2)}$ = $\mathcal{C}_A^2 \mathcal{C}_B^2$. All the other parameters are the same as Fig.~\ref{fig:site-dependent}.}
    \label{fig:firstfewsteps}
\end{figure*}

For a better understanding of Parrondo's paradox in the context of quantum walks, we plot the dynamics of the quantum walker over the first few steps in Fig.~\ref{fig:firstfewsteps}. These visualizations provide crucial insights into how the application of individual coin operators $\mathcal{C}_A$ and $\mathcal{C}_B$ and the combined operator influence the walker's probability distribution and overall trajectory. By examining the position probabilities at each step, we can observe the distinct interference patterns that emerge from the superposition of states induced by different coin operators. 

Plotting the expectation value of the position operator $\expval*{\hat{X}(t)}$ over these initial steps further illustrates the onset of Parrondo's paradox. While $\expval*{\hat{X}(t)}$ may decrease when a single coin operator is used repeatedly, the combined sequence results in a noticeable positive drift. These dynamics, captured through our plots, provide a visual confirmation of the paradox, showcasing the beneficial effects of combined strategies on the quantum walker's evolution from the very beginning.

\end{document}